\documentclass[11pt,english]{article}
\usepackage{babel}
\usepackage[dvips]{graphics}
\usepackage{epsfig}
\usepackage{latexsym}
\usepackage{amscd,amsmath,amssymb}
\usepackage[mathscr]{eucal}
\input xy
\xyoption{all}

\usepackage{amscd,amsmath,amssymb,bbm}

\numberwithin{equation}{section}
\newcommand{\id}[1]{\operatorname{id}_{#1}}

\newcommand{\D}[1]{\operatorname{d#1}}
\newcommand{\Tr}[1]{\operatorname{Tr #1}}
\newcommand{\diag}[1]{\operatorname{diag}(#1)}

\newcommand{\pbar}[1]{\partial_{\bar{z}^{\bar{#1}}}}

\newcommand{\F}[2]{F_{z^{#1}z^{#2}}}
\newcommand{\Fb}[2]{F_{\bar{z}^{\bar{#1}}\bar{z}^{\bar{#2}}}}
\newcommand{\Fbar}[2]{F_{z^{#1}\bar{z}^{\bar{#2}}}}
\newcommand{\f}[3]{F_{{#3}z^{#1}z^{#2}}}

\newcommand{\fbar}[3]{F_{{#3}z^{#1}\bar{z}^{\bar{#2}}}}
\newcommand{\sfrac}[2]{{\textstyle\frac{#1}{#2}}}

\begin{document}
\begin{titlepage}
\setcounter{page}{0}

\begin{flushright}
 hep-th/0304263\\ ITP--UH--01/03\\
\end{flushright}

\vskip 2.0cm

\begin{center}
 {\Large\bf Seiberg-Witten Monopole Equations\\
 on Noncommutative $\mathbbm{R}^4$}
\end{center}

\vspace*{0.3in}

\begin{center}
  {\large Alexander D. Popov$^{1,2}$, Armen G. Sergeev$^3$ and Martin
  Wolf$^{1,4}$}\\[0.2in]

  {\em $^1$Institut f\"ur Theoretische Physik, Universit\"at
  Hannover\\ Appelstra{\ss}e 2, 30167 Hannover, Germany\\[0.2in]

       $^2$Bogoliubov Laboratory of Theoretical Physics\\ JINR, 141980
       Dubna, Russia\\[0.2in]

       $^3$Steklov Mathematical Institute\\ Gubkina 8, 119991
       Moscow, Russia\\[0.2in]

       $^4$Institut f\"ur Theoretische Physik, Technische
       Universit\"at Dresden\\ 01062 Dresden, Germany}\\[0.2in]

       E-mail: {\ttfamily popov, wolf@itp.uni-hannover.de,
       sergeev@mi.ras.ru}
\end{center}

\vspace*{0.2in}

\begin{abstract}
\noindent It is well known that, due to vanishing theorems, there
are no nontrivial finite action solutions to the Abelian
Seiberg-Witten (SW) monopole equations on Euclidean
four-dimensional space $\mathbbm{R}^4$. We show that this is no
longer true for the noncommutative version of these equations,
i.e., on a noncommutative deformation $\mathbbm{R}^4_\theta$ of 
$\mathbbm{R}^4$ there exist smooth solutions to the SW equations 
having nonzero topological charge. We introduce
 action functionals for the noncommutative SW equations and 
construct explicit regular solutions. 
All our solutions have finite energy. We also suggest a possible 
interpretation of  
the obtained solutions as codimension four
vortex-like solitons representing $D(p-4)$- and 
$\overline{D(p-4)}$-branes in a $Dp$-$\overline{Dp}$ brane system in
type II superstring theory.

\end{abstract}

\end{titlepage}

\tableofcontents
\noindent\hrulefill

%---------------------------------------------------------------------

\section{Introduction}

The Seiberg-Witten (SW) monopole equations \cite{Witten:1994cg}
have been derived in the context of twisted $\mathcal{N}=2$
supersymmetric Yang-Mills (SYM) theory
\cite{Witten:1988ze,Seiberg:1994rs} in some limit
of the coupling constant. Another limit of this theory yields
the anti-self-dual Yang-Mills (ASDYM) equations. Namely, the
ASDYM equations correspond to the weak coupling limit while
the SW equations are related to the strong coupling regime
obtained by the S-dualization (see, e.g., \cite{Witten:1994cg,
Bilal:1995hc,Iga:2002cy,Marcolli} and references
therein). Note that the SW equations are associated with the
Abelian group $U(1)$ and have a compact moduli space while the
ASDYM equations, considered in Donaldson-Witten (DW) theory
\cite{Donaldson:1983,Witten:1988ze}, possess
the non-Abelian gauge group $SU(2)$ and a noncompact moduli
space. That is why SW theory is much easier to handle compared to DW
theory. A bridge between these theories is provided
by the non-Abelian SW equations (see, e.g., \cite{pid:1995,
Bradlow:1996,OkTel:1996,Labastida:1995zj,Marcolli}
and references therein) whose moduli space contains both DW and
SW moduli spaces as singular submanifolds.

It is well known that, due to a vanishing theorem of the
Lichnerowicz-Weitzenb\"ock type, there are no nontrivial finite
action solutions to the Abelian SW equations on Riemannian
four-manifolds with non-negative scalar curvature and, in particular,
on $\mathbbm{R}^4$ (cf. \cite{Witten:1994cg}). This assertion is
also true for lower-dimensional reductions of the SW equations,
i.e., these reductions also do not exhibit regular solutions on
$\mathbbm{R}^{n\leq 3}$ with a nonzero topological charge.
Nevertheless, one may construct nontrivial non-$L^2$ solutions,
as it has been done, e.g., in
\cite{Freund:1994ay,Nergiz:1996fu,Adam:2000tu,Legare:ri}.

Note that for the vanishing spinor field the SW equations
specialize to the Abelian ASDYM equations 
which
have no nontrivial regular solutions (instantons) on
$\mathbbm{R}^4$ either. However, Nekrasov and Schwarz
have demonstrated in \cite{Nekrasov:1998ss} that smooth Abelian
instanton solutions do exist on $\mathbbm{R}^4_\theta$ - a
noncommutative deformation of $\mathbbm{R}^4$ with constant
deformation parameters $\theta=(\theta^{\mu\nu})$. Moreover, they have
proven that noncommutativity resolves the singularities
of the instanton moduli space. In the present paper we observe a
similar phenomenon for the SW equations on
$\mathbbm{R}^4_\theta$ by constructing nontrivial regular
solutions to the noncommutative SW equations.

It is well known that the SW equations on K\"ahler surfaces are
similar to the vortex equations in two dimensions. Motivated by
this relation, we interpret regular (vortex-like) solutions to the
SW equations on $\mathbbm{R}^4_\theta$ as 
$D(p-4)$- and $\overline{D(p-4)}$-branes in a 
$Dp$-$\overline{Dp}$
brane-antibrane system in type II superstring theory. This
interpretation can also be extended to the commutative case of the
SW equations on K\"ahler surfaces.

The paper is organized as follows. In the next section we
formulate the SW equations on $\mathbbm{R}^4$ and fix our
notation. In section 3 we introduce the noncommutatively deformed
non-Abelian SW equations. We derive them from properly
deformed $U(2)$ self-duality type equations in eight dimensions
\cite{Corrigan:1982th} by a dimensional reduction to four
dimensions (cf.~\cite{Baulieu:1997jx}). The resulting $U_+(1)\times
U_-(1)$, $U_+(1)$ and $U_-(1)$ noncommutative SW equations can 
also be produced from appropriate action
functionals by using a Bogomolny type transformation. We point out
that the $U_+(1)\times U_-(1)$,
$U_+(1)$ and $U_-(1)$ noncommutative SW equations share
the same commutative limit. In section 4 we
present a number of regular solutions to the noncommutative SW
equations and discuss their $D$-brane interpretation in a string
theoretic context. In section 5 we conclude with a brief summary and
open problems. Finally, in the Appendix we perform the Bogomolny
type transformation for the noncommutative $U_+(1)\times U_-(1)$ 
SW action functional.

%---------------------------------------------------------------------

\section{SW monopole equations on $\mathbbm{R}^4$}

\paragraph{SW action functional.}
In this paper we consider the SW equations on the Euclidean space
$\mathbbm{R}^4$, provided with the standard metric
$g=(\delta_{\mu\nu})$, where $\mu, \nu,\ldots=1,\ldots,4$. The
(energy) functional $E=E(A,\Phi)$ for these equations has the
form (cf., e.g., \cite{Naber,Moore,Sergeev})
\begin{equation}\label{swa2}
  E(A,\Phi)=\int_{\mathbbm{R}^4}\D{ ^4x}\left\{|F_A|^2+
  |D_A\Phi|^2+\sfrac{1}{4}|\Phi|^4\right\}.
\end{equation}
Here $A\in\Omega^1(\mathbbm{R}^4,\mathfrak{u}(1))$ is a connection
one-form on $\mathbbm{R}^4$ with pure imaginary smooth
coefficients and $\Phi\in C^\infty(\mathbbm{R}^4,\mathbbm{C}^2)$ is
a Weyl spinor given by a smooth complex-valued 
vector function on $\mathbbm{R}^4$. We denote by
$F_A^+\in\Omega_+^2(\mathbbm{R}^4,\mathfrak{u}(1))$ the self-dual
part of the curvature $F_A$ of $A$ and by $D_A$ the covariant
derivative associated with $A$. Moreover, we use the abbreviation 
$|D_A\Phi|^2=D_\mu\phi_i(D_\mu\phi_i)^\dagger$
and set $|F_A|^2=\sfrac{1}{2}F_{\mu\nu}F_{\mu\nu}^\dagger$.

By exploiting a Bogomolny type formula, the energy functional can be
rewritten in the form
\begin{equation}\label{swaB1}
 E(A,\Phi)=SW(A,\Phi)-8\pi^2 Q,
\end{equation}
where
\begin{equation}\label{swa1}
 SW(A,\Phi)=\int_{\mathbbm{R}^4}\D{ ^4x}\left\{
 |\mathcal{D}_A\Phi|^2+2|F^+_A-
  \sigma^+(\Phi\otimes\Phi^\dagger)_0|^2\right\}
\end{equation}
is the SW action functional and
\begin{equation}\label{eq:tccc}
 Q=-\frac{1}{8\pi^2}\int_{\mathbbm{R}^4}F_A\wedge F_A
\end{equation}
is the topological charge. In the above formula (\ref{swa1}) we
denote by $\mathcal{D}_A$ the Dirac operator associated with $A$.
We also use the notation
\begin{equation}
 \sigma^+(\Phi\otimes\Phi^\dagger)_0:=
 \sigma^+(\Phi\otimes\Phi^\dagger-\sfrac{1}{2}|\Phi|^2\id{}),
\end{equation}
where
$$
\sigma^+\,:\,\text{Herm}_0(\mathbbm{C}^2)
\longrightarrow\Omega^2_+(\mathbbm{R}^4,\mathfrak{u}(1))
$$
is a map identifying the space $\text{Herm}_0(\mathbbm{C}^2)$ of
traceless Hermitian endomorphisms of $\mathbbm{C}^2$ with the
space $\Omega^2_+(\mathbbm{R}^4,\mathfrak{u}(1))$ of
imaginary-valued self-dual two-forms on $\mathbbm{R}^4$. The
inverse of this map is given by the Clifford multiplication by
two-forms (see, e.g., \cite{Naber,Moore,Sergeev}).

It is easy to see that the functionals (\ref{swa2}) and
(\ref{swa1}) are invariant under gauge transformations of the form
\begin{equation}\label{eq:gt}
 A\mapsto A+g^\dagger\D{}g\qquad\text{and}\qquad \Phi\mapsto
 g^\dagger\Phi,
\end{equation}
where $g\in C^\infty(\mathbbm{R}^4,U(1))$.

\paragraph{SW monopole equations.}
Since the functional $SW(A,\Phi)$ is positive semi-definite and
$Q$ is a topological term, the Bogomolny formula (\ref{swaB1})
implies that the lower bound of the energy $E(A,\Phi)$ is attained
on solutions to the equations
\begin{subequations}\label{swe1}
\begin{eqnarray}
 F^+_A &=& \sigma^+(\Phi\otimes\Phi^\dagger)_0,\\
 \mathcal{D}_A\Phi &=&0,
\end{eqnarray}
\end{subequations}
which are known as the SW monopole equations.
They are differential equations of first order and their
solutions, which minimize the energy functional $E(A,\Phi)$,
automatically satisfy the (second order) Euler-Lagrange equations
for the functionals $E(A,\Phi)$ and $SW(A,\Phi)$.

Writing $^t\Phi=(\phi_1,\phi_2)$,
one can see that the equations (\ref{swe1}) are equivalent to (cf.
\cite{Naber})
\begin{subequations}\label{swe2}
\begin{eqnarray}
 F_{12}+F_{34} &=& \sfrac{i}{2}
     (\phi_1\bar{\phi}_1-\phi_2\bar{\phi}_2),\notag\\
 F_{13}+F_{42} &=& -\sfrac{1}{2}
     (\phi_2\bar{\phi}_1-\phi_1\bar{\phi}_2),\\
 F_{14}+F_{23} &=& \sfrac{i}{2}
     (\phi_2\bar{\phi}_1+\phi_1\bar{\phi}_2)\notag
\end{eqnarray}
and
\begin{equation}\label{swe2-1}
 \begin{pmatrix} -D_4+iD_3 & D_2+iD_1
   \\ -D_2+iD_1 & -D_4-iD_3\end{pmatrix}
 \begin{pmatrix} \phi_1\\ \phi_2\end{pmatrix} =0.
\end{equation}
\end{subequations}

It is easy to prove that these equations have no nontrivial
solutions with finite action. Namely, we have the following theorem
(see, e.g.,~\cite{Witten:1994cg,Naber,Moore,Sergeev}):\\

\noindent
{\bf Theorem:}
 {\it Suppose $A\in\Omega^1(\mathbbm{R}^4,\mathfrak{u}(1))$ and
 $\Phi\in C^\infty(\mathbbm{R}^4,\mathbbm{C}^2)$ satisfy the
 equations (\ref{swe2}). Moreover, we assume that
 $\Phi\in L^2(\mathbbm{R}^4)$ and $E(A,\Phi)<\infty$. Then the only
 solution to (\ref{swe2}) is the trivial solution $(A,\Phi)=(0,0)$
 modulo the gauge transformations (\ref{eq:gt}).}\\

\noindent This theorem is also true for lower dimensional
reductions of the SW equations (defined on
$\mathbbm{R}^{n\leq3}$), i.e., these reductions do not exhibit
regular nontrivial solutions either. However, as we have already
mentioned in the Introduction, one can construct nontrivial
non-$L^2$ solutions 
\cite{Freund:1994ay,Nergiz:1996fu,Adam:2000tu,Legare:ri}.

\paragraph{Perturbed SW action functional and monopole equations.}
The gauge group action (\ref{eq:gt}) on the space of pairs
$(A,\Phi)$ is free, unless $\Phi\equiv 0$. In order to avoid
solutions of the form $(A,0)$, which may cause singularities in
the moduli space of solutions, we perturb the monopole equations
by adding an extra term to the first SW equation,
\begin{subequations}\label{swe3}
\begin{eqnarray}
 F^+_A+\chi^+ &=& \sigma^+(\Phi\otimes\Phi^\dagger)_0,\label{pe2}\\
 \mathcal{D}_A\Phi &=& 0,
\end{eqnarray}
\end{subequations}
where $\chi^+$ is the self-dual part of a two-form
$\chi\in\Omega^2(\mathbbm{R}^4,\mathfrak{u}(1))$ (perturbation).
Solutions to these equations minimize the functional
\begin{equation}\label{pswa}
 SW_\chi(A,\Phi)=\int_{\mathbbm{R}^4}\D{ ^4x}\left\{
 |\mathcal{D}_A\Phi|^2+2|F^+_A+\chi^+-
  \sigma^+(\Phi\otimes\Phi^\dagger)_0|^2\right\}.
\end{equation}
In components equation (\ref{pe2}) reads 
\begin{eqnarray}\label{eq:psweq}
 F_{12}+F_{34}+\chi_{12}+\chi_{34} &=&
 \sfrac{i}{2}(\phi_1\bar{\phi}_1-\phi_2\bar{\phi}_2),\notag\\
 F_{13}+F_{42}+\chi_{13}+\chi_{42} &=&
 -\sfrac{1}{2}(\phi_2\bar{\phi}_1-\phi_1\bar{\phi}_2),\\
 F_{14}+F_{23}+\chi_{14}+\chi_{23} &=&
 \sfrac{i}{2}(\phi_2\bar{\phi}_1+\phi_1\bar{\phi}_2).\notag
\end{eqnarray}

The SW action functional, as in the unperturbed case, is related to
an energy functional,
\begin{equation}\label{pswa2}
 E_\chi(A,\Phi)=\int_{\mathbbm{R}^4}\D{ ^4x}\left\{|F_A|^2+
  |D_A\Phi|^2+2|\chi^+-\sigma^+(\Phi\otimes\Phi^\dagger)_0|^2
  \right\},
\end{equation}
via a Bogomolny type formula,
\begin{equation}
 SW_\chi(A,\Phi)=E_\chi(A,\Phi)+16\pi^2\,K_\chi+8\pi^2 Q.
\end{equation}
The topological charge $Q$ is given, as before, by formula
(\ref{eq:tccc}) and the Chern-Simons type term $K_\chi$ is defined as
\begin{equation}\label{eq:cs-1}
 K_\chi=-\frac{1}{4\pi^2}\int_{\mathbbm{R}^4}\,F^+_A\wedge\chi^+.
\end{equation}

%---------------------------------------------------------------------

\section{SW monopole equations on $\mathbbm{R}^4_\theta$}

%---------------------------------------------------------------------

\subsection{Non-Abelian SW monopole equations}

\paragraph{Noncommutative Euclidean space $\mathbbm{R}^{2n}_\theta$.}
Let $\mathcal{A}(\mathbbm{R}^{2n})$ be the algebra of polynomial
functions on $\mathbbm{R}^{2n}$ (which is endowed with the canonical
metric
$\delta_{\alpha\beta}$) and $\theta=(\theta^{\alpha\beta})$ be a real
invertible skew-symmetric $2n\times 2n$ matrix with the inverse matrix
$\theta^{-1}=(\theta_{\alpha\beta})$ defined by
$\theta_{\alpha\gamma}\theta^{\gamma\beta}=\delta_{\alpha}^{\,\,\beta}$
for $\alpha,\beta,\ldots = 1,\ldots, 2n$.
Then the deformed algebra $\mathcal{A}_\theta(\mathbbm{R}^{2n})$ is defined
as
\begin{equation}\label{nces}
 \mathcal{A}_\theta(\mathbbm{R}^{2n}):=T(\mathbbm{R}^{2n})/
 \langle [x^\alpha,x^\beta]-i\theta^{\alpha\beta}\rangle_
{1\leq\alpha,\beta\leq 2n},
\end{equation}
where $T(\mathbbm{R}^{2n})$ is the tensor algebra of $\mathbbm{R}^{2n}$
and
$\langle [x^\alpha,x^\beta]-i\theta^{\alpha\beta}\rangle_
{1\leq\alpha,\beta\leq 2n}$
denotes the two-sided ideal generated by
$[x^\alpha,x^\beta]-i\theta^{\alpha\beta}\subset T(\mathbbm{R}^{2n})$.
{}For brevity we shall
denote $\mathcal{A}_\theta(\mathbbm{R}^{2n})$ simply by
$\mathbbm{R}^{2n}_\theta$ and call it the noncommutative Euclidean
$2n$-dimensional space.

One way to realize the noncommutative extension (\ref{nces})
of the algebra $\mathcal{A}(\mathbbm{R}^{2n})$ is by
deformation of the pointwise product between functions via the so-called
star (Moyal) product,
\begin{equation}\label{sp}
 (f\star g)(x):=f(x)\,\exp\left\{\sfrac{i}{2}
 \overleftarrow{\partial_{\alpha}}
 \theta^{\alpha\beta}\overrightarrow{\partial_{\beta}}\right\}\,g(x),
\end{equation}
where $f,g\in C^{\infty}(\mathbbm{R}^{2n},\mathbbm{C})$.
In particular, it follows from (\ref{sp}) that
\begin{equation}\label{cr1}
 [x^{\alpha},x^{\beta}]_{\star}:=
x^{\alpha}\star x^{\beta}-x^{\beta}\star x^{\alpha}
 =i\theta^{\alpha\beta}.
\end{equation}

For later convenience we introduce complex coordinates on
$\mathbbm{R}^{2n}\cong\mathbbm{C}^n$,
\begin{equation}\label{cc}
 z^a=x^{2a-1}+ix^{2a}\qquad\text{and}\qquad
 \bar{z}^{\bar{a}}=x^{2a-1}-ix^{2a},\qquad\text{for}\qquad a=1,\ldots,n,
\end{equation}
and derivatives
\begin{equation}
 \partial_{z^a}=\sfrac{1}{2}(\partial_{2a-1}-i\partial_{2a})
 \qquad\text{and}\qquad
 \pbar{a}=\sfrac{1}{2}(\partial_{2a-1}+i\partial_{2a}).
\end{equation}
Note that by an orthogonal change of coordinates one can always
transform $\theta^{\alpha\beta}$ to its canonical (Darboux) form
whose only nonzero components are $\theta^{2a-1,2a}$ with
$a=1,...,n$. Then the commutation relations $(\ref{cr1})$
translate to
\begin{equation}\label{cr2}
 [z^a,\bar{z}^{\bar{a}}]_\star=\theta^{a\bar{a}},
 \qquad\text{with}\qquad
 \theta^{a\bar{a}}=2\theta^{2a-1,2a},
\end{equation}
and all other commutators are equal to zero.

\paragraph{Self-duality type equations in eight dimensions.}
A standard way to obtain a noncommutative generalization of a
theory is to replace naively the ordinary commutative product
between field variables with the noncommutative star product.
However, it is well known that this method of translating a
commutative theory into a noncommutative one is not uniquely
defined when the matter fields are involved. For instance, the
scalar fields in noncommutative $U(1)$ gauge theory on
$\mathbbm{R}^{2n}$ can be regarded in three different ways, namely
as elements of a left module (over the algebra
$\mathcal{A}_\theta(\mathbbm{R}^{2n})$), or as elements of a right
module, or they can transform in the adjoint representation. For
this reason we propose deriving the noncommutative SW equations
from noncommutative self-duality type Yang-Mills (YM) equations in
eight dimensions, which are uniquely defined. Eventually, we will 
discover the equations corresponding to the above mentioned naive 
substitution rule by a formal reduction of more general equations.
In the commutative case a similar idea has been worked out by
the authors of \cite{Baulieu:1997jx}.

Let us consider pure $U(2)$ YM theory on
$\mathbbm{R}^8_\theta$. In star-product formulation the
components $F_{\alpha\beta}$ of the YM curvature $F_A$ read as
\begin{equation}
 F_{\alpha\beta}=\partial_\alpha A_\beta -\partial_\beta A_\alpha+
[A_\alpha,A_\beta]_\star
\end{equation}
and take values in $\mathfrak{u}(2)$. Here $\alpha,\beta,\ldots$, run
from $1$ to $8$.
Consider the generalized self-duality equations for YM fields in eight
dimensions{\footnote{In the commutative case these equations were
introduced in~\cite{Corrigan:1982th} and discussed, e.g., 
in~\cite{Ward,Fairlie}.}},
\begin{equation}\label{eq:sdoe}
 F_{\alpha\beta}=\sfrac{1}{2}T_{\alpha\beta\gamma\delta}F_{\gamma\delta},
\end{equation}
where the totally antisymmetric tensor $T_{\alpha\beta\gamma\delta}$
is determined by the octonionic structure constants $f_{ijk}$ as\footnote
{We use $f_{127}=f_{347}=f_{567}=f_{163}=f_{246}=
f_{253}=f_{154}=1$.}
\begin{equation}\label{eq:oct}
 T_{ijkl}=\sfrac{1}{6}\epsilon_{ijklmnq}f_{mnq}\qquad\text{and}\qquad
 T_{8ijk}=f_{ijk},\qquad\text{for}\qquad i,j,k,...=1,\ldots, 7.
\end{equation}
Note that the tensor $T_{\alpha\beta\gamma\delta}$ and therefore
the equations (\ref{eq:sdoe}) are invariant with respect to the
group $Spin(7)$ rather than $SO(8)$. In fact, it is impossible  to
construct a totally antisymmetric tensor of rank four in eight
dimensions which is invariant under $SO(8)$ rotations.

Using the definition (\ref{eq:oct}), we can write down the
generalized self-duality equations (\ref{eq:sdoe}) in components
as follows:
\begin{eqnarray}\label{eq:sdymo}
 F_{12}+F_{34}+F_{56}+F_{78} &=& 0,\notag\\
 F_{13}+F_{42}+F_{57}+F_{86} &=& 0,\notag\\
 F_{14}+F_{23}+F_{76}+F_{85} &=& 0,\notag\\
 F_{15}+F_{62}+F_{73}+F_{48} &=& 0,\\
 F_{16}+F_{25}+F_{38}+F_{47} &=& 0,\notag\\
 F_{17}+F_{82}+F_{35}+F_{64} &=& 0,\notag\\
 F_{18}+F_{27}+F_{63}+F_{54} &=& 0.\notag
\end{eqnarray}
With the help of (\ref{cc}) and the definitions
\begin{equation}
 A_{z^a}=\sfrac{1}{2}(A_{2a-1}-iA_{2a})\qquad\text{and}\qquad
 A_{\bar{z}^{\bar{a}}}=\sfrac{1}{2}(A_{2a-1}+iA_{2a}),\qquad\text{for}\qquad
 a=1,\ldots,4,
\end{equation}
we rewrite (\ref{eq:sdymo}) as
\begin{subequations}\label{eq:sdymococ}
\begin{eqnarray}
 \Fbar{1}{1}+\Fbar{2}{2}+\Fbar{3}{3}+\Fbar{4}{4} &=& 0,
 \label{eq:sdymoc1}\\
 \F{1}{2}+\Fb{3}{4} &=& 0,\label{eq:sdymoc2}\\
 \F{2}{4}-\Fb{1}{3} &=& 0,\label{eq:sdymoc3}\\
 \F{1}{4}+\Fb{2}{3} &=& 0.\label{eq:sdymoc4}
\end{eqnarray}
\end{subequations}
Note that $A_{\bar{z}^{\bar{a}}}=-A_{z^a}^\dagger$, since the
components $A_\alpha$ are skew-Hermitian.

\paragraph{Reduction to four dimensions.}
Following Baulieu et al.~\cite{Baulieu:1997jx}, we assume that the
gauge potential components $A_{z^a}$ for $a=1,\ldots,4$ do not
depend on the coordinates $z^3,z^4,\bar{z}^{\bar{3}},
\bar{z}^{\bar{4}}$ and define $^t\Psi:=(\Psi_1,\Psi_2)$ with
$\Psi_1:=A_{\bar{z}^{\bar{3}}}$ and $\Psi_2:=A_{z^4}$. Then the
equations (\ref{eq:sdymococ}) dimensionally reduce to
\begin{subequations}\label{eq:sdymococ-bla}
\begin{eqnarray}
 \Fbar{1}{1}+\Fbar{2}{2} &=& -([\Psi_1,\Psi_1^\dagger]_\star-
 [\Psi_2,\Psi_2^\dagger]_\star),\label{eq:sdymoc1-1}\\
 \F{1}{2} &=& [\Psi_1,\Psi_2^\dagger]_\star ,\label{eq:sdymoc2-1}\\
 D_{\bar{z}^{\bar{1}}}\Psi_1-D_{z^2}\Psi_2 &=& 0,\label{eq:sdymoc3-1}\\
 D_{\bar{z}^{\bar{2}}}\Psi_1+D_{z^1}\Psi_2 &=& 0,\label{eq:sdymoc4-1}
\end{eqnarray}
\end{subequations}
where the covariant derivative $D_{z^a}$ is defined by
$D_{z^a}\Psi=\partial_{z^a}\Psi+[A_{z^a},\Psi]_\star$. In the
commutative limit these equations coincide with a non-Abelian
generalization of the SW equations, considered in
\cite{Baulieu:1997jx}.
Note that the special case of these equations corresponding to
$\Psi_2\equiv0$ was discussed
in~\cite{Vafa:1994tf,Bonelli:1999it}.

Along with the unperturbed equations (\ref{eq:sdymococ-bla}) we
shall also consider the perturbed equations 
(cf. \cite{pid:1995,Bradlow:1996,OkTel:1996}). For that we introduce
a $\mathfrak{u}(2)$-valued two-form $\chi$ and add its self-dual
part $\chi^+$ to $F_A^+$ ,
\begin{subequations}\label{eq:per}
\begin{eqnarray}
 \Fbar{1}{1}+\Fbar{2}{2}+\chi_{z^1\bar{z}^{\bar{1}}}+
 \chi_{z^2\bar{z}^{\bar{2}}} &=& -([\Psi_1,\Psi_1^\dagger]_\star-
 [\Psi_2,\Psi_2^\dagger]_\star),\\
 \F{1}{2}+\chi_{z^1z^2} &=& [\Psi_1,\Psi_2^\dagger]_\star ,\\
 D_{\bar{z}^{\bar{1}}}\Psi_1-D_{z^2}\Psi_2 &=& 0,\\
 D_{\bar{z}^{\bar{2}}}\Psi_1+D_{z^1}\Psi_2 &=& 0.
\end{eqnarray}
\end{subequations}

%--------------------------------------------------------------------------

\subsection{Abelian SW monopole equations}

In order to get Abelian SW equations from the non-Abelian ones, we
shall consider solutions of a particular type given by a suitable
ansatz. This will reduce the gauge group $U(2)$ to $U(1)\times
U(1)$ and then, further down to $U(1)$.

\paragraph{Noncommutative $U_+(1)\times U_-(1)$ SW monopole equations.}
Let us consider the $U(1)\times U(1)$ subgroup of $U(2)$
with the generators
\begin{equation}
 \begin{pmatrix} i & 0\\ 0 & i\end{pmatrix}\qquad\text{and}\qquad
 \begin{pmatrix} i & 0\\ 0 & -i\end{pmatrix},
\end{equation}
and assume that the components of the gauge potential $A_{z^a}$
for $a=1,2$ take the form\footnote{If not stated differently,
$a,b,\ldots$, run always from $1$ to $2$ in the sequel.}
\begin{equation}\label{eq:defa}
 A_{z^a}:=\begin{pmatrix} A_{+z^a} & 0\\ 0 & A_{-z^a}\end{pmatrix},
 \qquad\text{with}\qquad A_{\pm z^a}=\sfrac{1}{2}(\mathcal{B}_{z^a}
 \pm\mathcal{A}_{z^a}),
\end{equation}
and $\mathcal{A}_{z^a},\mathcal{B}_{z^a}\in 
C^\infty(\mathbbm{R}^4,\text{i}\mathbbm{R}\otimes\mathbbm{C})$.
Furthermore, we restrict $\Psi_1$ and $\Psi_2$ to the form
\begin{equation}\label{eq:Psi}
 \Psi_{1,2}:=\begin{pmatrix} 0 & \sfrac{1}{\sqrt{8}}\phi_{1,2}\\
       0 & 0\end{pmatrix},
\end{equation}
where $\phi_{1,2}\in C^\infty(\mathbbm{R}^4,\mathbbm{C})$.

Substituting (\ref{eq:defa}) and (\ref{eq:Psi}) into the equations
(\ref{eq:sdymococ-bla}), after a straightforward calculation
we obtain the equations
\begin{subequations}\label{eq:swbf}
\begin{eqnarray}
 \fbar{1}{1}{+}+\fbar{2}{2}{+}=-\sfrac{1}{8}
 (\phi_1\star\phi_1^\dagger-\phi_2\star\phi_2^\dagger)
 \quad &\text{and}&\quad \f{1}{2}{+}=
  \sfrac{1}{8}\phi_1\star\phi_2^\dagger,\label{eq:swcurv1}\\
 \fbar{1}{1}{-}+\fbar{2}{2}{-}=\sfrac{1}{8}
    (\phi_1^\dagger\star\phi_1-\phi_2^\dagger\star\phi_2)
 \quad &\text{and}&\quad
 \f{1}{2}{-}=-\sfrac{1}{8}\phi_2^\dagger\star\phi_1,
 \label{eq:swcurv2}
\end{eqnarray}
\end{subequations}
as well as
\begin{equation}\label{eq:swbfd}
 D_{\bar{z}^{\bar{1}}}\phi_1-D_{z^2}\phi_2 = 0
 \qquad\text{and}\qquad
 D_{\bar{z}^{\bar{2}}}\phi_1+ D_{z^1}\phi_2 =0,
\end{equation}
where we have used the definition
\begin{equation}\label{eq:cd}
 D_{z^a}\phi := \partial_{z^a}\phi+A_{+z^a}\star\phi-\phi\star
 A_{-z^a}.
\end{equation}
Here $\fbar{a}{a}{\pm}$ are the components of the curvature
associated with $A_{\pm z^a}$, i.e.,
\begin{equation}
 \fbar{a}{a}{\pm}=\partial_{z^a}A_{\pm\bar{z}^{\bar{a}}}
 -\partial_{\bar{z}^{\bar{a}}}A_{\pm z^a}+[A_{\pm z^a},
 A_{\pm\bar{z}^{\bar{a}}}]_\star.
\end{equation}

Analogously, by assuming that $\chi=\diag{\chi_+,\chi_-}$ with
$\chi_\pm\in\Omega^2(\mathbbm{R}^4,\text{i}\mathbbm{R})$ in
(\ref{eq:per}), we obtain the perturbed equations
\begin{subequations}\label{eq:swbfp}
\begin{eqnarray}
 \fbar{1}{1}{+}+\fbar{2}{2}{+}+\chi_{+z^1\bar{z}^{\bar{1}}}
 +\chi_{+z^2\bar{z}^{\bar{2}}}
 &=&-\sfrac{1}{8}(\phi_1\star\phi_1^\dagger-\phi_2\star\phi_2^\dagger),
 \label{eq:swbfp1}\\
 \f{1}{2}{+}+\chi_{+z^1z^2}&=&\sfrac{1}{8}\phi_1\star\phi_2^\dagger,
 \label{eq:swbfp2}\\
 \fbar{1}{2}{-}+\fbar{2}{2}{-}+\chi_{-z^1\bar{z}^{\bar{1}}}
 +\chi_{-z^2\bar{z}^{\bar{2}}}\label{eq:swbfp3}
 &=&\sfrac{1}{8}(\phi_1^\dagger\star\phi_1-
 \phi_2^\dagger\star\phi_2),\\
 \f{1}{2}{-}+\chi_{-z^1z^2}&=&-\sfrac{1}{8}\phi_2^\dagger\star
 \phi_1.\label{eq:swbfp4}
\end{eqnarray}
\end{subequations}
We consider the equations (\ref{eq:swbf}), (\ref{eq:swbfd})
and (\ref{eq:swbfp}), (\ref{eq:swbfd}) as a
noncommutative extension of the unperturbed and perturbed
Abelian SW equations, respectively. Since there are two gauge 
potentials in the equations, we call them the noncommutative 
unperturbed and perturbed 
$U_+(1)\times U_-(1)$ SW equations.

It remains to find out what kind of gauge transformations leave
the $U_+(1)\times U_-(1)$ SW equations invariant. It is obvious
from the explicit form of these equations that $\phi_1$ and
$\phi_2$ are in the bi-fundamental representation of $U_+(1)\times
U_-(1)$. Hence, the equations (\ref{eq:swbf}), (\ref{eq:swbfd})
and (\ref{eq:swbfp}) are invariant under gauge transformations of
the form
\begin{equation}\label{eq:gtbf}
 A_{\pm}\mapsto g_\pm^\dagger\star A_{\pm}\star g_\pm+
 g_\pm^\dagger\star\D{} g_\pm,\qquad
 \chi_\pm\mapsto g_\pm^\dagger\star\chi_\pm\star g_\pm
 \qquad\text{and}\qquad \Phi\mapsto g_+^\dagger\star\Phi\star g_-,
\end{equation}
where $g_\pm\in C^\infty(\mathbbm{R}^4,U_\pm(1))$ and
$^t\Phi=(\phi_1,\phi_2)$.

In the commutative limit
the covariant derivative (\ref{eq:cd}) turns into
\begin{equation}
 D_{z^a}\phi = \partial_{z^a}\phi+(A_{+z^a}-A_{-z^a})\phi=
 \partial_{z^a}\phi+\mathcal{A}_{z^a}\phi,
\end{equation}
i.e., the gauge potential $\mathcal{B}$ disappears from the 
equations (\ref{eq:swbfd}). In other words, one copy of $U(1)$
decouples from $U(2)$ and the matter field $\Phi$
interacts only with the $SU(2)$ part. 
Hence, in the commutative case $\Phi$ is charged with respect
to the diagonal $U(1)$ subgroup of $U_+(1)\times U_-(1)$ 
corresponding to the gauge potential $\mathcal{A}=A_+-A_-$.
Furthermore, the commutator in the expression for the
(Abelian) curvature vanishes and hence as a
corollary we have 
$\fbar{a}{a}{\pm}=\sfrac{1}{2}(\fbar{a}{a}{\mathcal{B}\,}
\pm\fbar{a}{a}{\mathcal{A}\,})$.
In the commutative limit one may
choose the perturbations $\chi_\pm$ so that
\begin{equation}
 \chi_{\pm z^a\bar{z}^{\bar{a}}}=-\sfrac{1}{2}
   F_{\mathcal{B}\,z^a\bar{z}^{\bar{a}}}\pm 
 \sfrac{1}{2}\chi_{z^a\bar{z}^{\bar{a}}}
 \qquad \text{and} \qquad
 \chi_{\pm z^1z^2}=-\sfrac{1}{2}F_{\mathcal{B}\,z^1z^2}\pm 
 \sfrac{1}{2}\chi_{z^1z^2},
\end{equation}
where $\chi\in\Omega^2(\mathbbm{R}^4,\mathfrak{u}(1))$
is some other perturbation. Then
from equations (\ref{eq:swbfp})
we obtain\footnote{Note that in the case of complex
valued-functions `$^\dagger$' goes to `$^-$' in the commutative limit.}
\begin{equation}\label{eq:blabla}
  F_{\mathcal{A}\,z^1\bar{z}^{\bar{1}}}+
  F_{\mathcal{A}\,z^2\bar{z}^{\bar{2}}}+\chi_{z^1\bar{z}^{\bar{1}}}+
  \chi_{z^2\bar{z}^{\bar{2}}}=-\sfrac{1}{4}
  (\phi_1\bar{\phi}_1-\phi_2\bar{\phi}_2)
  \quad \text{and}\quad F_{\mathcal{A}\,z^1z^2}+
  \chi_{z^1z^2}=\sfrac{1}{4}\phi_1\bar{\phi}_2.
\end{equation}
Thus, we recover the perturbed SW equations (\ref{swe2-1}) and
(\ref{eq:psweq}) (written in complex coordinates). Of course, the
choice $\chi\equiv 0$ corresponds to the unperturbed equations.

\paragraph{Remark.}
Consider the unperturbed equations (\ref{eq:swbf}) and (\ref{eq:swbfd}).
In the commutative limit we arrive at the standard unperturbed SW
equations for configurations ($\mathcal{A},\Phi$) plus the Abelian
ASDYM equations for $\mathcal{B}$, i.e.,
\begin{equation}
  F_{\mathcal{B}\,z^1\bar{z}^{\bar{1}}}+
  F_{\mathcal{B}\,z^2\bar{z}^{\bar{2}}}=0
  \qquad \text{and}\qquad
  F_{\mathcal{B}\,z^1z^2}= 0.
\end{equation}
Taking the trivial solution $\mathcal{B}=0$ (recall that there are
no Abelian instantons on $\mathbbm{R}^4$) we remain with the
standard unperturbed SW equations. More generally, any pure gauge
configuration for $\mathcal{B}$ will do the same. The
noncommutative version of the latter statement is, however, nontrivial. 
If we choose $\mathcal{B}_{z^a}$ in (\ref{eq:defa}) of the
form
\begin{equation}\label{eq:decb}
 \mathcal{B}_{z^a}=\sfrac{1}{2}
 (b^\dagger\star\partial_{z^a}b-b\star\partial_{z^a}
  b^\dagger),
\end{equation}
with $b\in C^\infty(\mathbbm{R}^4,U(1))$, then it will correspond
in the commutative limit to a pure gauge configuration but,
of course, it is
not pure gauge in the noncommutative case. Only in the
commutative limit the curvature $F_{\mathcal{B}}$ disappears and
we arrive at the unperturbed SW equations (\ref{swe2}).

\paragraph{Noncommutative $U_\pm(1)$ SW monopole equations.}
In equations (\ref{eq:swbf}) - (\ref{eq:gtbf}) the field $\Phi$ 
is regarded as an element of a $\mathbbm{R}_\theta^4$-bimodule
transforming in the bi-fundamental representation of the gauge group
$U_+(1)\times U_-(1)$. However, in the noncommutative setup the matter
field $\Phi$ can also be thought of either as an element of a right
$\mathbbm{R}_\theta^4$-module (the $U_+(1)$ case) or as an element
of a left $\mathbbm{R}^4_\theta$-module (the $U_-(1)$ case). These
two cases can easily be read off the equations (\ref{eq:swbfp1}), 
(\ref{eq:swbfp2}) and (\ref{eq:swbfp3}), (\ref{eq:swbfp4}), respectively.  
Namely, consider the equations\footnote{Formally, these equations
can be obtained from (\ref{eq:swbfd}) - (\ref{eq:swbfp}) by choosing 
$\mathcal{B}_{z^a}=\mathcal{A}_{z^a}$ (i.e., $A_{-z^a}=0$), 
taking $\chi_\pm$ such that
$\chi_{+z^a\bar{z}^{\bar{a}}}=\chi_{z^a\bar{z}^{\bar{a}}}$,
$\chi_{+z^1z^2}=\chi_{z^1z^2}$,
$\chi_{-z^1\bar{z}^{\bar{1}}}+ \chi_{-z^2\bar{z}^{\bar{2}}}=\sfrac{1}{8}
    (\phi_1^\dagger\star\phi_1-\phi_2^\dagger\star\phi_2)$
and
 $\chi_{-z^1z^2}=-\sfrac{1}{8}\phi_2^\dagger\star\phi_1$,
where $\chi\in\Omega^2(\mathbbm{R}^4,\text{i}\mathbbm{R})$, and rescaling
$\Phi\mapsto\sqrt{2}\Phi$.}
\begin{subequations}\label{eq:uplus}
\begin{equation}
  F_{z^1\bar{z}^{\bar{1}}}+ 
  F_{z^2\bar{z}^{\bar{2}}}+
  \chi_{z^1\bar{z}^{\bar{1}}}+\chi_{z^2\bar{z}^{\bar{2}}}
 =-\sfrac{1}{4}
 (\phi_1\star\phi_1^\dagger-\phi_2\star\phi_2^\dagger) 
 \quad \text{and} \quad  F_{z^1z^2}+\chi_{z^1z^2}=
  \sfrac{1}{4}\phi_1\star\phi_2^\dagger,
\end{equation}
as well as
\begin{equation}
 D_{\bar{z}^{\bar{1}}}\phi_1-D_{z^2}\phi_2 = 0
 \qquad\text{and}\qquad
 D_{\bar{z}^{\bar{2}}}\phi_1+D_{z^1}\phi_2=0,
\end{equation}
where $\chi\in\Omega^2(\mathbbm{R}^4,\text{i}\mathbbm{R})$ and
the (right) covariant derivative reads
\begin{equation}\label{eq:uplus-cd}
 D_{z^a}\phi = \partial_{z^a}\phi+\mathcal{A}_{z^a}\star\phi.
\end{equation}
\end{subequations}
Note that the curvature $F_{\mathcal{A}}$ is now computed from 
$\mathcal{A}$, i.e.,
$F_{\mu\nu}=\partial_\mu\mathcal{A}_\nu-\partial_\nu\mathcal{A}_\mu+
[\mathcal{A}_\mu,\mathcal{A}_\nu]_\star$.

Similarly, we may introduce the equations\footnote{These equations
can formally be obtained from (\ref{eq:swbfd}) - (\ref{eq:swbfp}) 
by choosing $\mathcal{B}_{z^a}=-\mathcal{A}_{z^a}$ (i.e., $A_{+z^a}=0$) 
and putting
 $\chi_{+z^1\bar{z}^{\bar{1}}}+\chi_{+z^2\bar{z}^{\bar{2}}}=
 -\sfrac{1}{8}
   (\phi_1\star\phi_1^\dagger-\phi_2\star\phi_2^\dagger)$,
 $ \chi_{+z^1z^2}=\sfrac{1}{8}\phi_1\star\phi_2^\dagger$,
 $\chi_{-z^a\bar{z}^{\bar{a}}}=\chi_{z^a\bar{z}^{\bar{a}}}$ and
$\chi_{-z^1z^2}=\chi_{z^1z^2}$,
where $\chi\in\Omega^2(\mathbbm{R}^4,\text{i}\mathbbm{R})$. Again
we should rescale $\Phi\mapsto\sqrt{2}\Phi$.}
\begin{subequations}\label{eq:uminus}
\begin{equation}
  F_{z^1\bar{z}^{\bar{1}}}+ 
  F_{z^2\bar{z}^{\bar{2}}}+
  \chi_{z^1\bar{z}^{\bar{1}}}+\chi_{z^2\bar{z}^{\bar{2}}}
 =\sfrac{1}{4}
 (\phi_1^\dagger\star\phi_1-\phi_2^\dagger\star\phi_2) 
 \quad \text{and} \quad  F_{z^1z^2}+\chi_{z^1z^2}=
  -\sfrac{1}{4}\phi_2^\dagger\star\phi_1
\end{equation}
and
\begin{equation}
 D_{\bar{z}^{\bar{1}}}\phi_1-D_{z^2}\phi_2 = 0
 \qquad\text{and}\qquad
 D_{\bar{z}^{\bar{2}}}\phi_1+D_{z^1}\phi_2 =0,
\end{equation}
where $\chi\in\Omega^2(\mathbbm{R}^4,\text{i}\mathbbm{R})$ 
and
\begin{equation}
 D_{z^a}\phi = \partial_{z^a}\phi-\phi\star\mathcal{B}_{z^a}
\end{equation}
\end{subequations}
is the (left) derivative. Now the curvature $F_{\mathcal{B}}$ 
is associated with the gauge potential 
$\mathcal{B}$, i.e.,
$F_{\mu\nu}=\partial_\mu\mathcal{B}_\nu-\partial_\nu\mathcal{B}_\mu+
[\mathcal{B}_\mu,\mathcal{B}_\nu]_\star$.

We shall call (\ref{eq:uplus}) and (\ref{eq:uminus}) 
the perturbed noncommutative $U_+(1)$ and $U_-(1)$ 
SW equations. 
Obviously, the unperturbed equations appear for $\chi\equiv 0$.
Note that the systems (\ref{eq:uplus}) and (\ref{eq:uminus})  
are totally equivalent and
the only difference between them is an artifact of noncommutativity.
The commutative limits of both cases are, of course, identical and
produce (\ref{swe3}). Moreover, in the commutative 
case the gauge transformations (\ref{eq:gtbf}) reduce
to the standard ones, i.e., one may choose the identity either for
$g_-$ or for $g_+$. 
 
%---------------------------------------------------------------------

\subsection{Operator form of the Abelian SW monopole equations}

\paragraph{Weyl transform.}
Due to the nonlocal nature of the star product, explicit
calculations might be quite tedious. It is therefore convenient to
pass over to the operator formalism via the Weyl ordering
$\mathcal{W}$ given by
\begin{subequations}
\begin{eqnarray}
 \mathcal{W}\,:\,\tilde{f}(k)\mapsto\hat{f}(\hat{x}) &=&
 \frac{1}{(2\pi)^{2n}}\int_{\mathbbm{R}^{2n}}\D{ ^{2n}k}\,
 \tilde{f}(k)\,e^{ik\hat{x}},\\
 \mathcal{W}^{-1}\,:\,\hat{f}(\hat{x})\mapsto\tilde{f}(k) &=&
 |\text{Pf}(2\pi\theta)|\Tr{}\{e^{-ik\hat{x}}\hat{f}(\hat{x})\},
\end{eqnarray}
\end{subequations}
where  $\tilde{f}(k)$ stands for the
Fourier transform of $f(x)\in\mathcal{S}(\mathbbm{R}^{2n})$,
\begin{equation}
 f(x)\mapsto\tilde{f}(k)=\int_{\mathbbm{R}^{2n}}\D{ ^{2n}x}\,f(x)e^{-ikx}.
\end{equation}
Here $\mathcal{S}(\mathbbm{R}^{2n})$ is the Schwartz space of fast
decreasing functions\footnote{However, in later considerations we
shall make suitable choices for $f$ which are weaker.} on
$\mathbbm{R}^{2n}$, `$\Tr{}$' denotes the trace in the operator
representation of the noncommutative algebra and
`$\text{Pf}(2\pi\theta)$' is the Pfaffian of
$(2\pi\theta^{\alpha\beta})$. Also, in these equations $kx$ is a
shorthand notation for $k_{\alpha}x^{\alpha}$. One can verify (see,
e.g.,~\cite{Douglas:2001ba}) that the following relations are true:
\begin{equation}
 \mathcal{W}\,:\,f\star g\mapsto \hat{f}\hat{g}\qquad\text{and}\qquad
 \int_{\mathbbm{R}^{2n}}\D{ ^{2n}x}\,f=|\text{Pf}(2\pi\theta)|\Tr{}\hat{f}.
\end{equation}

We may regard the coordinates $\hat{x}^\alpha$ as operators which
act on some Fock space $\mathcal{H}$, specified in section
\ref{sec:solpsw}, and satisfy the commutation relations
$[\hat{x}^\alpha,\hat{x}^\beta]= i\theta^{\alpha\beta}$. With a
proper choice of coordinates the parameters $\theta^{\alpha\beta}$
will have the canonical form (\ref{cr2}). For the complex
coordinates $\hat{z}^a$, also considered as operators in
$\mathcal{H}$, we then get
\begin{equation}\label{cr3}
 [\hat{z}^a,\hat{z}^b]=0\qquad\text{and}\qquad
 [\hat{z}^a,\hat{\bar{z}}^{\bar{a}}]=\theta^{a\bar{a}},
 \qquad\text{for}\qquad
 a,b=1,\ldots,n.
\end{equation}
A straightforward calculation shows that coordinate
derivatives are now inner derivations of this algebra, i.e.,
\begin{equation}\label{der1}
 \hat{\partial}_{z^a}\hat{f}=
    \theta_{a\bar{a}}[\hat{\bar{z}}^{\bar{a}},\hat{f}]
 \qquad\text{and}\qquad
 \hat{\partial}_{\bar{z}^{\bar{a}}}\hat{f}=
    \theta_{\bar{a}a}[\hat{z}^a,\hat{f}].
\end{equation}
In the operator formulation, the perturbed noncommutative
$U_+(1)\times U_-(1)$ SW equations (\ref{eq:swbfd}) and
(\ref{eq:swbfp}) retain their form,
\begin{subequations}\label{eq:swbfpo}
\begin{eqnarray}
 \hat{F}_{+z^1\bar{z}^{\bar{1}}}+\hat{F}_{+z^2\bar{z}^{\bar{2}}}+
 \hat{\chi}_{+z^1\bar{z}^{\bar{1}}}+\hat{\chi}_{+z^2\bar{z}^{\bar{2}}}
 &=&-\sfrac{1}{8}(\hat{\phi}_1\hat{\phi}_1^\dagger
 -\hat{\phi}_2\hat{\phi}_2^\dagger),
 \label{eq:swbfpo1}\\
 \hat{F}_{+z^1z^2}+\hat{\chi}_{+z^1z^2}&=&\sfrac{1}{8}\hat{\phi}_1
 \hat{\phi}_2^\dagger,\\
 \hat{F}_{-z^1\bar{z}^{\bar{1}}}+\hat{F}_{-z^2\bar{z}^{\bar{2}}}+
 \hat{\chi}_{-z^1\bar{z}^{\bar{1}}}+\hat{\chi}_{-z^2\bar{z}^{\bar{2}}}
 &=&\sfrac{1}{8}(\hat{\phi}_1^\dagger\hat{\phi}_1-
 \hat{\phi}_2^\dagger\hat{\phi}_2),\\
 \hat{F}_{-z^1z^2}+\hat{\chi}_{-z^1z^2}&=&-\sfrac{1}{8}
 \hat{\phi}_2^\dagger\hat{\phi}_1,\label{eq:swbfpo4}
\end{eqnarray}
and
\begin{equation}\label{eq:swbfdo}
 \hat{D}_{\bar{z}^{\bar{1}}}\hat{\phi}_1-\hat{D}_{z^2}\hat{\phi}_2
 = 0
 \qquad\text{and}\qquad
 \hat{D}_{\bar{z}^{\bar{2}}}\hat{\phi}_1+\hat{D}_{z^1}\hat{\phi}_2
 =0,
\end{equation}
\end{subequations}
where
\begin{equation}\label{eq:cdo}
 \hat{D}_{z^a}\phi = \hat{\partial}_{z^a}\hat{\phi}+
 \hat{A}_{+z^a}\hat{\phi}-\hat{\phi}\hat{A}_{-z^a}.
\end{equation}

\noindent In order to simplify our notation, from now on we omit
the hats over the operators.

\paragraph{$U_+(1)\times U_-(1)$ SW action functional.} 
Having introduced the $U_+(1)\times U_-(1)$ 
noncommutative extension of the SW equations (\ref{eq:swbfpo}), 
we shall define an appropriate action functional. For
this purpose we switch back to real coordinates.

Let $^t\Phi=(\phi_1,\phi_2)$ and 
${^t\Phi}^*:=(\phi_2^\dagger,-\phi_1^\dagger)$.\footnote{Note 
that `$*$' is nothing but a spinor conjugation.} 
Then the noncommutative deformation of the action functional 
(\ref{pswa}) will have the form
\begin{equation}\label{ncpswa}
\begin{split}
 SW_\chi(A_+,A_-,\Phi;\theta)=&\sfrac{1}{2}|\text{Pf}(2\pi\theta)|\Tr{}
  \left\{|\mathcal{D}_{A_+,A_-}\Phi|^2+
  |(\mathcal{D}_{A_+,A_-}\Phi)^\dagger|^2\right. \\
  &\left.\hspace{-.5in}+\,\,8|F^+_{A_+}+\chi^+_+-\sigma^+
  (\Phi\otimes\Phi^\dagger)_0|^2+8|F^+_{A_-}+\chi^+_--\sigma^+
  (\Phi^*\otimes(\Phi^*)^\dagger)_0|^2\right\},
\end{split}
\end{equation}
where 
\begin{equation}
 |\psi|^2:=\psi_1\psi_1^\dagger+\psi_2\psi_2^\dagger
 \qquad\text{and}\qquad
 |\psi^\dagger|^2:=\psi_1^\dagger\psi_1+\psi_2^\dagger\psi_2,
\end{equation}
for any $^t\psi=(\psi_1,\psi_2)$. Here, $\mathcal{D}_{A_+,A_-}$ 
denotes the Dirac operator
depending on the two gauge potentials $A_+$ and $A_-$, i.e.,
\begin{equation}
 \mathcal{D}_{A_+,A_-}\Phi=
  \begin{pmatrix} -D_4+iD_3 & D_2+iD_1\\
                  -D_2+iD_1 & -D_4-iD_3
  \end{pmatrix}
  \begin{pmatrix} \phi_1\\ \phi_2\end{pmatrix},
\end{equation} 
where the covariant derivatives $D_\mu$ are given by (\ref{eq:cdo}).
Note that the prefactors in (\ref{ncpswa}) are adjusted in such a 
way that we recover (\ref{pswa}) in the commutative limit.
It is not difficult to see that the (perturbed) $U_+(1)\times U_-(1)$
SW equations following from (\ref{ncpswa}) are given by the equations
\begin{equation}
 \mathcal{D}_{A_+,A_-}\Phi=0,\qquad
 F^+_{A_+}+\chi^+_+=\sigma^+(\Phi\otimes\Phi^\dagger)_0
 \qquad\text{and}\qquad
 F^+_{A_-}+\chi^+_-=\sigma^+(\Phi^*\otimes(\Phi^*)^\dagger)_0,
\end{equation}
whose solutions minimize the action functional (\ref{ncpswa}).
In components these equations coincide with (\ref{eq:swbfpo}).

Assuming that $\phi_{1,2}$ and $F_{\pm\mu\nu}$ are of proper
trace-class,
e.g., $|\Tr{}\phi_{1,2}|<\infty$ and $|\Tr{}F_{\pm\mu\nu}|<\infty$, 
we can show that
\begin{equation}\label{bog}
 SW_\chi(A_+,A_-,\Phi;\theta)=E_\chi(A_+,A_-,\Phi;\theta)
 +16\pi^2\,K_\chi-
 |\text{Pf}(2\pi\theta)|\Tr{}\mathcal{T},
\end{equation} 
where the functionals $E_\chi$ and $K_\chi$ are 
given by 
\begin{equation}\label{eq:energy}
\begin{split}
 E_\chi(A_+,A_-,\Phi;\theta)=&|\text{Pf}(2\pi\theta)|
 \Tr{}\left\{2|F_{A_+}|^2+2|F_{A_-}|^2+
 \sfrac{1}{2}|D_{A_+,A_-}\Phi|^2+
 \sfrac{1}{2}|(D_{A_+,A_-}\Phi)^\dagger|^2\right.\\
 &\left.\qquad+\,\,4|\chi^+_+-\sigma^+(\Phi\otimes\Phi^\dagger)_0|^2+
 4|\chi^+_--\sigma^+(\Phi^*\otimes(\Phi^*)^\dagger)_0|^2 \right\}
\end{split}
\end{equation}
and
\begin{equation}\label{eq:kchi}
 K_\chi=-\frac{1}{8\pi^2}|\text{Pf}(2\pi\theta)|
 \Tr{}\sum_{i=\pm}\,\{F^+_{i\mu\nu},\chi^+_{i\mu\nu}\}.
\end{equation}
The topological term $\mathcal{T}$ reads
\begin{subequations}
\begin{equation}\label{eq:topo}
 \mathcal{T} = \sum_{i=\pm}\left(
 F_{i\mu\nu}\,{*F_{i\mu\nu}}+\nabla_{i\mu}\mathcal{J}_{i\mu}
 \right).
\end{equation}
Here
\begin{equation}\label{nabla1}
 \nabla_{\pm\mu}\,\cdot\,:=
 \partial_\mu\,\cdot\,+[A_{\pm\mu},\,\cdot\,\,],
\end{equation}
\end{subequations} 
`$*$' denotes the Hodge operator and $\{A,B\}:=AB+BA$. 
The currents $\mathcal{J}_{\pm\mu}$ 
depend in a particular fashion on the fields $\phi_1$ and $\phi_2$, 
their derivatives and on the gauge potentials $A_{\pm\mu}$. 
The explicit derivation of (\ref{bog}) and the expressions for the 
currents $\mathcal{J}_{\pm\mu}$ are given in the Appendix. 
Note that similarly to the commutative case
the functionals $E_\chi+16\pi^2K_\chi$ and $SW_\chi$ yield the same
equations of motion. The integration of (\ref{eq:topo}) yields the 
topological charge 
\begin{equation}\label{tc} 
 Q=-\frac{1}{8\pi^2}|\text{Pf}(2\pi\theta)|\Tr{}\mathcal{T},
\end{equation}
for the considered field configuration 
$(A_+,A_-,\Phi)$ on $\mathbbm{R}^4_\theta$.
 
\paragraph{Generalized coupled vortex equations.}
Let us put one component of $\Phi$ to zero, e.g. consider the
case  $^t\Phi=(\phi_1,\phi_2)=:(\phi,0)$. \footnote{Note that in the 
case of K\"ahler manifolds ($\mathbbm{R}^4$ is trivially K\"ahler) the
field $\phi$ can be regarded as a scalar \cite{Moore,Sergeev}.}
Moreover, we choose
\begin{equation}
 \chi_{\pm z^1z^2}=0\qquad\text{and}\qquad
 \chi_{\pm z^1\bar{z}^{\bar{1}}}+\chi_{\pm z^2\bar{z}^{\bar{2}}}
 =\mp\sfrac{1}{8}\,v_\pm,
\end{equation}
where $v_{\pm}$ are some Hermitian operators acting on 
$\mathcal{H}$. Then the energy functional (\ref{eq:energy}) turns 
into
\begin{subequations}\label{eq:e-k-chi}
\begin{equation}\label{eq:echi}
\begin{split}
 E_\chi(A_+,A_-,\Phi;\theta)=& |\text{Pf}(2\pi\theta)|\Tr{}
 \left\{F_{+\mu\nu}F_{+\mu\nu}^\dagger
 +F_{-\mu\nu}F_{-\mu\nu}^\dagger+\sfrac{1}{2}D_\mu\phi
 (D_\mu\phi)^\dagger+\sfrac{1}{2}(D_\mu\phi)^\dagger 
 D_\mu\phi\right.\\
 &\left.\hspace*{1in}+\,\,\sfrac{1}{8}(v_+-\phi\phi^\dagger)^2
 +\sfrac{1}{8}(v_--\phi^\dagger\phi)^2\right\},
\end{split}
\end{equation} 
and $K_\chi$ is given by
\begin{equation}\label{eq:kchi-1}
 K_\chi=-\frac{i}{32\pi^2}|\text{Pf}(2\pi\theta)|
 \Tr{}\left\{\{F_{+12}+F_{+34},v_+\}-\{F_{-12}+F_{-34},v_-\}\right\}.
\end{equation}
\end{subequations}
Also the currents $\mathcal{J}_{\pm\mu}$ have a fairly simple 
form\footnote{Cf. the Appendix.},
\begin{subequations}\label{eq:currents}
\begin{eqnarray}
 \mathcal{J}_{+\mu}&=&\sfrac{i}{4}(\epsilon_{\mu\nu12}+\epsilon_{\mu\nu34})
           (\phi(D_\nu\phi)^\dagger-(D_\nu\phi)\phi^\dagger),\label{rc1}\\
 \mathcal{J}_{-\mu}&=&-\sfrac{i}{4}(\epsilon_{\mu\nu12}+\epsilon_{\mu\nu34})
           (\phi^\dagger(D_\nu\phi)-(D_\nu\phi)^\dagger\phi),\label{rc1-2}
\end{eqnarray}
\end{subequations}
where $\epsilon_{\mu\nu\lambda\sigma}$ is the Levi-Civita
symbol with $\epsilon_{1234}=1$. 

For our choices of $\Phi$ and $\chi_\pm$ the
perturbed $U_+(1)\times U_-(1)$ SW equations in complex
coordinates read 
\begin{subequations}\label{eq:uglugl}
\begin{eqnarray}
 F_{+z^1\bar{z}^{\bar{1}}}+F_{+z^2\bar{z}^{\bar{2}}}
 =\sfrac{1}{8}(v_+-\phi\phi^\dagger)
 \qquad &\text{and}& \qquad 
 F_{+z^1z^2}=0,\label{eq:blabla1}\\
 F_{-z^1\bar{z}^{\bar{1}}}+F_{-z^2\bar{z}^{\bar{2}}}
 =-\sfrac{1}{8}(v_--\phi^\dagger\phi)
 \qquad &\text{and}&\qquad
 F_{-z^1z^2}=0,\label{eq:blabla2}\\
 D_{\bar{z}^{\bar{1}}}\phi=0
 \qquad &\text{and}& \qquad
 D_{\bar{z}^{\bar{2}}}\phi=0.\label{eq:blabla0}
\end{eqnarray}
\end{subequations}
In the commutative case these equations were considered, e.g., in 
\cite{Bradlow:1996}.

\paragraph{$U_\pm(1)$ SW action functionals.} 
Having introduced the $U_+(1)\times U_-(1)$ SW functionals, we are
now interested in  proper functionals for the $U_\pm(1)$ cases
(\ref{eq:uplus}) and (\ref{eq:uminus}).
Let us first discuss the perturbed $U_+(1)$ SW equations.
In this case the SW action functional takes the following form:
\begin{equation}\label{ncpswa-uplus}
 SW_\chi(\mathcal{A},\Phi;\theta)=|\text{Pf}(2\pi\theta)|\Tr{}
  \{|\mathcal{D}_{\mathcal{A}}\Phi|^2+2|F^+_{\mathcal{A}}+\chi^+-
  \sigma^+(\Phi\otimes\Phi^\dagger)_0|^2\}.
\end{equation}
Note that now the Dirac operator depends only on $\mathcal{A}$ and 
the covariant derivatives are given by (\ref{eq:uplus-cd}).
Also here the prefactors have been chosen such that the correct 
commutative limit will be obtained.
As before, the functional $SW_\chi$ may be rewritten as
\begin{equation}
 SW_\chi(\mathcal{A},\Phi;\theta)=E_\chi(\mathcal{A},\Phi;\theta)+
 16\pi^2K_\chi-|\text{Pf}(2\pi\theta)|\Tr{}\mathcal{T},
\end{equation} 
where $E_\chi$ turns out to be   
\begin{equation}
 E_\chi(\mathcal{A},\Phi;\theta)=|\text{Pf}(2\pi\theta)|
 \Tr{}\left\{|F_{\mathcal{A}}|^2+|D_{\mathcal{A}}\Phi|^2
 +2|\chi^+-\sigma^+(\Phi\otimes\Phi^\dagger)_0|^2\right\}.
\end{equation}
The Chern-Simons term $K_\chi$ reads as
\begin{equation}
 K_\chi=-\frac{1}{16\pi^2}|\text{Pf}(2\pi\theta)|
 \Tr{}\,\{F^+_{\mu\nu},\chi^+_{\mu\nu}\},
\end{equation}
and the topological term $\mathcal{T}$ is given by
\begin{equation}
 \mathcal{T} = \sfrac{1}{4}\{F_{\mu\nu},{*F_{\mu\nu}}\}+
 \nabla_{\mu}\mathcal{J}_{\mu},\qquad\text{with}\qquad
 \nabla_\mu\mathcal{J}_\mu=\partial_\mu\mathcal{J}_\mu+
 [\mathcal{A}_\mu,\mathcal{J}_\mu],
\end{equation}
implying the charge
\begin{equation}\label{eq:tc-uplus}
 Q=-\frac{1}{8\pi^2}|\text{Pf}(2\pi\theta)|\Tr{}\mathcal{T}.
\end{equation}

Again all equations simplify essentially if one chooses
 $^t\Phi=(\phi_1,\phi_2)=(\phi,0)$, $\chi_{z^1z^2}=0$ and
$\chi_{z^1\bar{z}^{\bar{1}}}+\chi_{z^2\bar{z}^{\bar{2}}}
=-\sfrac{1}{4}v$, where $v$ is some Hermitian 
operator. Then $E_\chi$ and $K_\chi$ are given by
\begin{subequations}
\begin{equation}
 E_\chi(\mathcal{A},\Phi;\theta)=|\text{Pf}(2\pi\theta)|
 \Tr{}\left\{\sfrac{1}{2}F_{\mu\nu}F_{\mu\nu}^\dagger
 +D_\mu\phi(D_\mu\phi)^\dagger
 +\sfrac{1}{4}(v-\phi\phi^\dagger)^2\right\}
\end{equation}
and
\begin{equation}
 K_\chi=-\frac{i}{32\pi^2}|\text{Pf}(2\pi\theta)|
 \Tr{}\,\{F_{12}+F_{34},v\}.
\end{equation}
\end{subequations}
The current $\mathcal{J}_\mu$ reduces in this case to
\begin{equation}\label{eq:current-uplus}
 \mathcal{J}_{\mu}=\sfrac{i}{2}(\epsilon_{\mu\nu12}+\epsilon_{\mu\nu34})
           (\phi(D_\nu\phi)^\dagger-(D_\nu\phi)\phi^\dagger).
\end{equation}
Finally, the perturbed $U_+(1)$ SW equations read 
\begin{subequations}\label{eq:uglugl-uplus}
\begin{eqnarray}
 F_{z^1\bar{z}^{\bar{1}}}+F_{z^2\bar{z}^{\bar{2}}}
 =\sfrac{1}{4}(v-\phi\phi^\dagger)
  \qquad &\text{and}&\qquad 
 F_{z^1z^2}=0,\label{eq:blabla1-uplus}\\
 D_{\bar{z}^{\bar{1}}}\phi=0
 \qquad &\text{and}& \qquad
 D_{\bar{z}^{\bar{2}}}\phi=0.\label{eq:blabla0-uplus}
\end{eqnarray}
\end{subequations}

In the commutative case these $v$-vortex equations in four dimensions
were considered, e.g., in \cite{Bradlow:1996}.
Note that in a similar manner one can write down the functionals 
corresponding to the
$U_-(1)$ case (\ref{eq:uminus}). Since they look essentially the same
as the above-introduced $U_+(1)$ functionals we refrain from writing 
down their explicit form.

%--------------------------------------------------------------------

\section{Particular solutions}\label{sec:solpsw}

\subsection{Operator realization}

The form of the Heisenberg algebra type commutation relations 
(\ref{cr3}) suggests that the algebra $\mathbbm{R}^4_\theta$ may 
be represented by
a pair of harmonic oscillators in the two-oscillator Fock space
$\mathcal{H}\cong\mathcal{H}_1\otimes
\mathcal{H}_2\cong\bigoplus_{n_1,n_2}\mathbbm{C}\,|n_1,n_2\rangle$.
We introduce, as usual, annihilation and creation operators
$\{c_a,c_a^\dagger\}_{a=1,2}$, satisfying $[c_a,c_a^\dagger]=1$.
They act on $\mathcal{H}$ and are defined by the relations
\begin{subequations}\label{eq:cao}
\begin{eqnarray}
 c_1|n_1,n_2\rangle &=& \sqrt{n_1}\,|n_1-1,n_2\rangle
    \qquad\text{and}\qquad
 c_1^\dagger|n_1,n_2\rangle = \sqrt{n_1+1}\,|n_1+1,n_2\rangle,\\
 c_2|n_1,n_2\rangle &=& \sqrt{n_2}\,|n_1,n_2-1\rangle
    \qquad\text{and}\qquad
 c_2^\dagger|n_1,n_2\rangle = \sqrt{n_2+1}\,|n_1,n_2+1\rangle,
\end{eqnarray}
\end{subequations}
where $\{|n_1,n_2\rangle\,|\,n_1,n_2\in\mathbbm{N}_0\}$ form an
orthonormal basis in $\mathcal{H}$. The commutation relations
(\ref{cr3}) imply that the operators $\{c_a,c_a^\dagger\}_{a=1,2}$
have the form
\begin{subequations}\label{eq:cao-z}
   \begin{align}
    c_1:=\sfrac{\hat{z}^1}{\sqrt{\theta^{1\bar{1}}}}
         \sfrac{1+\text{sgn}(\theta^{1\bar{1}})}{2}
        +\sfrac{\hat{\bar{z}}^{\bar{1}}}{\sqrt{\theta^{\bar{1}1}}}
        \sfrac{1-\text{sgn}(\theta^{1\bar{1}})}{2}
    &\quad\text{and}\quad
    c_1^\dagger:=\sfrac{\hat{\bar{z}}^{\bar{1}}}
        {\sqrt{\theta^{1\bar{1}}}}\sfrac{1+\text{sgn}
        (\theta^{1\bar{1}})}{2}
        +\sfrac{\hat{z}^1}{\sqrt{\theta^{\bar{1}1}}}
        \sfrac{1-\text{sgn}(\theta^{1\bar{1}})}{2},\\
    c_2:=\sfrac{\hat{z}^2}{\sqrt{\theta^{2\bar{2}}}}
        \sfrac{1+\text{sgn}(\theta^{2\bar{2}})}{2}
        +\sfrac{\hat{\bar{z}}^{\bar{2}}}
        {\sqrt{\theta^{\bar{2}2}}}
        \sfrac{1-\text{sgn}(\theta^{2\bar{2}})}{2}
    &\quad\text{and}\quad
    c_2^\dagger:=\sfrac{\hat{\bar{z}}^{\bar{2}}}
        {\sqrt{\theta^{2\bar{2}}}}\sfrac{1+\text{sgn}
        (\theta^{2\bar{2}})}{2}
        +\sfrac{\hat{z}^2}{\sqrt{\theta^{\bar{2}2}}}
        \sfrac{1-\text{sgn}(\theta^{2\bar{2}})}{2}.
   \end{align}
\end{subequations}

We introduce so-called shift operators $S^{(a)}$ acting on the
Fock spaces $\mathcal{H}_{a}$. They are partially isometric
operators sending the Fock space $\mathcal{H}_a$ to its subspace
$(\mathbbm{1}^{(a)}\!\!-\!\!P_0^{(a)})\mathcal{H}_a$, where we denote by
$P^{(a)}_0=|0\rangle_a\langle 0|_a$
 the orthogonal projector onto
the ground state of $\mathcal{H}_a$ and
$\mathbbm{1}^{(a)}\!\!-\!\!P_0^{(a)}$ is the complement projector. Then
\begin{equation}
 S^{(a)}\,:\,\mathcal{H}_a\rightarrow (\mathbbm{1}^{(a)}\!\!
 -\!\!P_0^{(a)})\mathcal{H}_a,\quad\text{with}
 \quad S^{(a)\dagger}S^{(a)}=\mathbbm{1}^{(a)}\quad\text{and}\quad
 S^{(a)}S^{(a)\dagger} = \mathbbm{1}^{(a)}\!\!-\!\!P_0^{(a)}.
\end{equation}
The operator $S^{(a)}$ may be given by the explicit formula
\begin{equation}
 S^{(a)}=\sum_{n\geq 0}|n+1\rangle_a\langle n|_a.
\end{equation}
We will sometimes drop the index `$a$' on the state $|n\rangle_a$ in
the following if the meaning is clear from the context.

The next step is to construct a shift operator $S$
on $\mathcal{H}\cong\mathcal{H}_1\otimes \mathcal{H}_2$ such that
\begin{equation}
 S\,:\,\mathcal{H}\rightarrow (\mathbbm{1}-P_0)\mathcal{H},
 \qquad\text{with}\qquad P_0=|0,0\rangle\langle 0,0|,
\end{equation}
and $S^\dagger S=\mathbbm{1}$ and $SS^\dagger=\mathbbm{1}-P_0$.
A naive idea to take simply the tensor product $S^{(1)}\otimes
S^{(2)}$ does not work. One possible realization of the required
operator $S$ is given by the formula \cite{Furuuchi:2000vc}
\begin{equation}\label{shift1}
 S=\mathbbm{1}+\sum_{n\geq 0}\,(|n+1\rangle\langle n|-
 |n\rangle\langle n|) \otimes P^{(2)}_0=\mathbbm{1}+
 (S^{(1)}\!\!-\!\!\mathbbm{1}^{(1)})\otimes P_0^{(2)},
\end{equation}
but there are also other realizations (see,
e.g.,~\cite{Lechtenfeld:2001ie}).

For later convenience we introduce the operators
\begin{equation}\label{soa}
 X_{\pm z^a}:=A_{\pm z^a}+\theta_{a\bar{a}}\bar{z}^{\bar{a}}
   \qquad\text{and}\qquad
 X_{\pm\bar{z}^{\bar{a}}}:=
  A_{\pm\bar{z}^{\bar{a}}}+\theta_{\bar{a}a}z^a,
 \qquad\text{for}\qquad a=1,2.
\end{equation}
Then a short calculation of the YM curvature yields
\begin{subequations}\label{eq:curvature}
\begin{equation}
 \f{1}{2}{\pm}= [X_{\pm z^1},X_{\pm z^2}],\qquad
 \fbar{1}{2}{\pm}=[X_{\pm z^1},X_{\pm\bar{z}^{\bar{2}}}],\qquad
 \fbar{2}{1}{\pm}=[X_{\pm z^2},X_{\pm\bar{z}^{\bar{1}}}],\qquad
\end{equation}
\begin{equation}
 \fbar{1}{1}{\pm} = [X_{\pm z^1},X_{\pm\bar{z}^{\bar{1}}}]+\theta_{1\bar{1}},
 \qquad
 \fbar{2}{2}{\pm} = [X_{\pm z^2},X_{\pm\bar{z}^{\bar{2}}}]+\theta_{2\bar{2}},
\end{equation}
\end{subequations}
and the covariant derivatives become
\begin{equation}
 D_{z^a}\phi=X_{+z^a}\phi-\phi X_{-z^a}
 \qquad\text{and}\qquad
 D_{\bar{z}^{\bar{a}}}\phi
 =X_{+\bar{z}^{\bar{a}}}\phi-\phi X_{-\bar{z}^{\bar{a}}}.
\end{equation}

%---------------------------------------------------------------------

\subsection{Solutions to the perturbed SW equations}

\paragraph{$U_+(1)\times U_-(1)$ SW monopole equations.}
Let us consider the equations (\ref{eq:uglugl}) rewritten in operator 
form. For the operators (\ref{soa}) we take (cf., e.g., 
\cite{Gross:2000ss,Harvey:2000jb})
\begin{equation}\label{eq:anx}
 X_{\pm z^a}=\theta_{a\bar{a}}S^N\bar{z}^{\bar{a}}(S^\dagger)^N+
 \sum_{n=0}^{N-1}\lambda_{a,n}|n\rangle\langle n|\otimes P_0^{(2)},
\end{equation}
where the shift operator $S$ is given by (\ref{shift1}),
$\lambda_{a,n}\in\mathbbm{C}$ and $N\in\mathbbm{N}$. Then the 
commutator $[X_{\pm z^a},X_{\pm\bar{z}^{\bar{a}}}]$ is readily
computed to be
\begin{equation}\label{eq:commutator}
 [X_{\pm z^a},X_{\pm\bar{z}^{\bar{a}}}]=-\theta_{a\bar{a}}
 (\mathbbm{1}-\mathcal{P}_N),
\end{equation}
where $\mathcal{P}_N$ is given by 
\begin{equation}
 \mathcal{P}_N:=\sum_{n=0}^{N-1}|n\rangle\langle n|\otimes P_0^{(2)}.
\end{equation}
It easy to see that the second equations of (\ref{eq:blabla1}) and
(\ref{eq:blabla2}) are trivially satisfied. Choosing $v_-\equiv 0$ 
and
\begin{equation}
 \phi=\sqrt{8(\theta_{1\bar{1}}+\theta_{2\bar{2}})}\,
 \mathcal{P}_N,
\end{equation}
we can solve (\ref{eq:blabla0}) and the first equation of 
(\ref{eq:blabla2}) consistently, while the first equation of 
(\ref{eq:blabla1}) implies that
\begin{equation}
 v_+=16(\theta_{1\bar{1}}+\theta_{2\bar{2}})\mathcal{P}_N.
\end{equation}
Note that the moduli $\lambda_{a,n}$ in (\ref{eq:anx}) can be
interpreted as position parameters (see, e.g., 
\cite{Gross:2000ss,Hashimoto:2001pc,Hamanaka:2003cm}).

Obviously, the components (\ref{eq:curvature}) of the
curvature and the 
field $\phi$ are of proper trace-class. Moreover,
it can be easily checked that $\phi$ is covariantly constant, i.e., 
along with $D_{\bar{z}^{\bar{a}}}\phi=0$, required by the SW
equations, we also have $D_{z^a}\phi=0$. This means
that the currents $\mathcal{J}_{\pm\mu}$ (\ref{eq:currents}) 
vanish identically. 
Thus, it is straightforward to evaluate the 
topological charge (\ref{tc}) for this configuration. What we 
find for the topological term (\ref{eq:topo}) is
\begin{equation}
 \mathcal{T}=-8\frac{1}{\theta^{12}\theta^{34}}\mathcal{P}_N.
\end{equation}
Using  $|\text{Pf}(2\pi\theta)|=4\pi^2|\theta^{12}\theta^{34}|$, 
we get immediately a charge\footnote{Note that the definition
of the charge (\ref{tc}) differs by a factor of $2$ in comparison
with the standard one.}, 
\begin{equation}
 Q=4\epsilon_1\epsilon_2N,\qquad\text{with}\qquad 
 \epsilon_1:=\frac{|\theta^{12}|}{\theta^{12}}
 \qquad\text{and}\qquad \epsilon_2
 :=\frac{|\theta^{34}|}{\theta^{34}}. 
\end{equation}
Note that $K_\chi$ given by (\ref{eq:kchi-1})
is also finite. Therefore, the considered
field configuration has finite energy 
\begin{equation}\label{eq:energy-chi}
 E_\chi=32\pi^2 f(\theta) N\quad\text{with}\quad f(\theta)
 :=|\theta^{12}\theta^{34}|\left[\left(\frac{1}{\theta^{12}}
 \right)^2+\frac{1}{\theta^{12}\theta^{34}}+\left(
 \frac{1}{\theta^{34}}\right)^2\right].
\end{equation}

Let us now consider a slight generalization of the ansatz 
(\ref{eq:anx}). We take first
\begin{subequations}\label{eq:ax-ap}
\begin{equation}\label{eq:ax}
 X_{\pm z^a}=\theta_{a\bar{a}}S^{N_\pm}\bar{z}^{\bar{a}}
 (S^\dagger)^{N_\pm}+
 \sum_{n=0}^{N_\pm-1}\lambda_{a,n}^{\pm}|n\rangle\langle n|\otimes 
 P_0^{(2)},
\end{equation}
and find that $[X_{\pm z^a},X_{\pm\bar{z}^{\bar{a}}}]=
-\theta_{a\bar{a}}(\mathbbm{1}-\mathcal{P}_{N_\pm})$. Second, choosing
\begin{equation}\label{eq:ap}
 \phi=\sqrt{8(\theta_{1\bar{1}}+\theta_{2\bar{2}})}\,S^{N_+}
 (S^\dagger)^{N_-}
\end{equation}
\end{subequations}
and the perturbations $v_\pm$ such that
\begin{equation}
v_+=8(\theta_{1\bar{1}}+\theta_{2\bar{2}})\qquad\text{and}\qquad
v_- =8(\theta_{1\bar{1}}+\theta_{2\bar{2}})(1-2\mathcal{P}_{N_-}),
\end{equation}
one can easily show that our equations are solved
consistently. Again, all our operators are of proper trace-class for
$N_+\neq N_-$. In the case of $N_+=N_-$ 
one encounters a slight subtlety since
the field $\phi$ is not of trace-class contrary to our assumption. 
However, potentially dangerous terms, like $\phi F_{-12}\phi^\dagger$ 
for instance\footnote{Cf. the Appendix.}, which occured 
in (\ref{eq:energy}), are obviously zero when $N_+=N_-$. 
Moreover, the field $\phi$ is covariantly constant, as one can readily 
check.
Therefore, the currents $\mathcal{J}_{\pm\mu}$ (\ref{eq:currents}) 
are identically zero.  The topological term 
(\ref{eq:topo}) for this configuration thus reads as
\begin{equation}
 \mathcal{T}=-4\frac{1}{\theta^{12}\theta^{34}}(\mathcal{P}_{N_+}+
 \mathcal{P}_{N_-}),
\end{equation}
which produces a topological charge$^{11}$ 
$Q=2\epsilon_1\epsilon_2(N_++N_-)$.  
The functional $E_\chi$ for this solution
computes to 
\begin{equation}
 E_\chi=16\pi^2f(\theta)(N_++N_-), 
\end{equation}
where $f(\theta)$ is given by (\ref{eq:energy-chi}).

\paragraph{$U_+(1)$ SW monopole equations.}
Consider now the equations (\ref{eq:uglugl-uplus}) and choose
the ansatz
\begin{equation}\label{eq:ax-uplus} 
 X_{z^a}=\theta_{a\bar{a}}S^N\bar{z}^{\bar{a}}(S^\dagger)^N+
 \sum_{n=0}^{N-1}\lambda_{a,n}|n\rangle\langle n|\otimes 
 P_0^{(2)}\qquad\text{and}\qquad\phi=\gamma S^N, 
\end{equation}
with $\lambda_{a,n}\in\mathbbm{C}$ and $\gamma\in\mathbbm{R}$. 
Then the first equation of (\ref{eq:blabla1-uplus}) can be solved by 
$\theta_{1\bar{1}}+\theta_{2\bar{2}}=\sfrac{1}{4}\gamma^2=
\sfrac{1}{4}v$,
while the other two equations are trivially
satisfied. The only nonvanishing components of 
the curvature are $\fbar{a}{a}{}=\theta_{a\bar{a}}\mathcal{P}_N$. Note that 
the expression for $\phi$ is covariantly constant which  
implies the vanishing of the current (\ref{eq:current-uplus}). 
The charge (\ref{eq:tc-uplus}) 
is equal to $Q=\epsilon_1\epsilon_2N$ and the energy $E_\chi$ is 
$8\pi^2f(\theta)N$,
where $f(\theta)$ is given by (\ref{eq:energy-chi}).
 
Consider again the equations (\ref{eq:uglugl-uplus}). Besides the
shift type solution for $\phi$ we can also find a 
projector type solution. Namely, we choose
$\phi=\gamma P_0$ with $\gamma\in\mathbbm{R}$, 
\begin{equation}
 v=(4(\theta_{1\bar{1}}+\theta_{2\bar{2}})+\gamma^2)P_0,
\end{equation}
and the $X_{z^a}$'s as previously. 
With this choice the first two equations of 
(\ref{eq:blabla1-uplus}) are trivially satisfied, while 
(\ref{eq:blabla0-uplus}) yields the condition
$P_0z^a=0$. Hence, we have to set $\theta^{a\bar{a}}<0$.
Again, the only nonvanishing components of the curvature are 
$\fbar{a}{a}{}=\theta_{a\bar{a}}P_0$.
Moreover, we have 
\begin{equation}
 D_1\phi=-iD_2\phi=-\sqrt{\theta_{1\bar{1}}}\gamma
                   |0,0\rangle\langle 1,0|
 \qquad\text{and}\qquad
  D_3\phi=-iD_4\phi=-\sqrt{\theta_{2\bar{2}}}\gamma
                   |0,0\rangle\langle 0,1|,
\end{equation}
which imply the vanishing of current the $\mathcal{J}_{\mu}$ 
(\ref{eq:current-uplus}). The 
topological charge  for this configuration
is $Q=\epsilon_1\epsilon_2$.

%---------------------------------------------------------------------

\subsection{Solutions to the unperturbed SW equations}
\label{sec:solsw}

In this section we shall consider solutions to the unperturbed
SW equations. We concentrate on the case of the
unperturbed $U_+(1)$ equations, i.e., (\ref{eq:uglugl-uplus}) with 
$v\equiv 0$, as an illustrative example.

\paragraph{$U_+(1)$ SW monopole equations with $\Phi\equiv 0$.} 
In this case we obtain the noncommutative Abelian ASDYM equations. 
Solutions to these equations, i.e., noncommutative Abelian instantons, 
have been known for quite some time~\cite{Nekrasov:1998ss}. 
However, for the sake of completeness we briefly review their 
construction. Note that in the case $\Phi\equiv 0$ the current 
$\mathcal{J}_{\mu}$ is identically zero. 

In terms of the operators (\ref{soa}) the Abelian ASDYM equations read 
\begin{equation}\label{asdym}
 [X_{z^1},X_{\bar{z}^{\bar{1}}}]+ [X_{z^2},X_{\bar{z}^{\bar{2}}}]
 +\theta_{1\bar{1}}+\theta_{2\bar{2}}=0\qquad\text{and}\qquad
 [X_{z^1},X_{z^2}]=0.
\end{equation}   
Again we consider an ansatz for $X_{z^a}$ of the form 
(\ref{eq:ax-uplus}).
This ansatz yields solutions to the equations 
(\ref{asdym}) if the deformation tensor $\theta^{\mu\nu}$
is anti-self-dual which trivially follows from (\ref{eq:ax-uplus}).
The only nonvanishing 
components of the curvature are $\fbar{a}{a}{}=\theta_{a\bar{a}}
\mathcal{P}_N$ implying that the charge (\ref{eq:tc-uplus}) is equal
to $Q=-N$. 
The moduli $\lambda_{a,n}$ entering the solution are
position parameters showing the location of
the noncommutative instantons \cite{Furuuchi:2000dx} 
(see also \cite{Hamanaka:2003cm} for a recent review).

It is also possible to consider
the case of ASDYM equations on a self-dual 
background for which 
$\theta^{1\bar{1}}=\theta^{2\bar{2}}=:\theta$. 
Let us assume that $\theta>0$, which leads to the definitions 
$c_1=z^1/\sqrt{\theta}$ and
$c_2=z^2/\sqrt{\theta}$. We
choose the ansatz~\cite{Kraus:2001xt,Nekrasov:2002kc}
\begin{equation} 
 X_{z^1}=-\sfrac{1}{\theta} S^\dagger \bar{z}^{\bar{1}}f(N)S
 =-\sfrac{1}{\sqrt{\theta}}S^\dagger c_1^\dagger
 f(N)S\quad\text{and}\quad
 X_{z^2}=-\sfrac{1}{\theta} S^\dagger \bar{z}^{\bar{2}}f(N)S
 =-\sfrac{1}{\sqrt{\theta}}S^\dagger c_2^\dagger f(N)S,
\end{equation}
and assume that
$f(N)|0\rangle=f(0)|0\rangle =0$, where 
$N:=N_1+N_2:=c_1^\dagger c_1+c_2^\dagger c_2$.
Then the equation 
$[X_{z^1},X_{z^2}]=0$ is
trivially satisfied. A short calculation yields for $f(N)$
the result~\cite{Kraus:2001xt,Nekrasov:2002kc}
\begin{equation}
 f^2(N)=\frac{N(N+3)}{(N+1)(N+2)}.
\end{equation}
The nonvanishing components of the curvature in this case are
\begin{subequations}
\begin{eqnarray}
 \fbar{a}{b}{}&=&\sfrac{1}{\theta} 
 S^\dagger c_a^\dagger(f^2(N+1)-f^2(N))c_bS,\\
 \fbar{a}{a}{}&=&\sfrac{1}{\theta} 
 S^\dagger [(N_a+1)f^2(N)-N_af^2(N-1)-1]S.
\end{eqnarray}
\end{subequations} 
Using these expressions, we compute the topological charge
(\ref{eq:tc-uplus}) to be $-1$. A straightforward extension of the above 
ansatz allows
one to construct multi-instanton configurations, as well
\cite{Kraus:2001xt,Nekrasov:2002kc}.

\paragraph{Fock spaces with indefinite norm.}
Let us now discuss the case when $\Phi$ does not vanish
identically. For that we relax the condition of positivity
of the norm of the Fock spaces $\mathcal{H}_{1,2}$ and assume instead
that at least one of them has an indefinite norm
(cf.~\cite{Bognar}). For instance, we can introduce a collection
of the creation and annihilation operators
$\{c_a,c_a^\dagger\}_{a=1,2}$, satisfying $[c_1,c_1^\dagger]=1$,
$[c_2,c_2^\dagger]=-1$, and defined by the relations
\begin{subequations}
\begin{eqnarray}
 c_1|n_1,n_2\rangle &=& \sqrt{n_1}\,|n_1-1,n_2\rangle
    \qquad\,\,\,\,\,\text{and}\qquad
 c_1^\dagger|n_1,n_2\rangle = \sqrt{n_1+1}\,|n_1+1,n_2\rangle,\\
 c_2|n_1,n_2\rangle &=& -\sqrt{n_2}\,|n_1,n_2-1\rangle
    \qquad\text{and}\qquad
 c_2^\dagger|n_1,n_2\rangle = \sqrt{n_2+1}\,|n_1,n_2+1\rangle,
\end{eqnarray}
\end{subequations}
substituting (\ref{eq:cao}). The normalization condition is modified 
to $\langle
n_1,n_2|m_1,m_2\rangle=(-1)^{n_2}\delta_{n_1m_1} \delta_{n_2m_2}$.
The identity operator is given by
\begin{equation}\label{eq:unew}
    \mathbbm{1}=\sum_{n_1,n_2}(-1)^{n_2}
    |n_1,n_2\rangle\langle n_1,n_2|.
\end{equation}
Moreover, we have to redefine the relations between
$\{c_a,c_a^\dagger\}_{a=1,2}$ and
$\{z^a,\bar{z}^{\bar{a}}\}_{a=1,2}$ so that
   \begin{subequations}\label{eq:crea}
   \begin{align}
    c_1:=\sfrac{\hat{z}^1}{\sqrt{\theta^{1\bar{1}}}}
         \sfrac{1+\text{sgn}(\theta^{1\bar{1}})}{2}
        +\sfrac{\hat{\bar{z}}^{\bar{1}}}
        {\sqrt{\theta^{\bar{1}1}}}
        \sfrac{1-\text{sgn}(\theta^{1\bar{1}})}{2}
    &\quad\text{and}\quad
    c_1^\dagger:=\sfrac{\hat{\bar{z}}^{\bar{1}}}
        {\sqrt{\theta^{1\bar{1}}}}\sfrac{1+\text{sgn}
        (\theta^{1\bar{1}})}{2}
        +\sfrac{\hat{z}^1}{\sqrt{\theta^{\bar{1}1}}}
        \sfrac{1-\text{sgn}(\theta^{1\bar{1}})}{2},\\
    c_2:=\sfrac{\hat{\bar{z}}^{\bar{2}}}
        {\sqrt{\theta^{2\bar{2}}}}\sfrac{1+\text{sgn}
        (\theta^{2\bar{2}})}{2}
        +\sfrac{\hat{z}^2}{\sqrt{\theta^{\bar{2}2}}}
        \sfrac{1-\text{sgn}(\theta^{2\bar{2}})}{2}
    &\quad\text{and}\quad
    c_2^\dagger:=\sfrac{\hat{z}^2}
        {\sqrt{\theta^{2\bar{2}}}}
        \sfrac{1+\text{sgn}(\theta^{2\bar{2}})}{2}
        +\sfrac{\hat{\bar{z}}^{\bar{2}}}
        {\sqrt{\theta^{\bar{2}2}}}
        \sfrac{1-\text{sgn}(\theta^{2\bar{2}})}{2}.
   \end{align}
   \end{subequations}
The definition of the shift operator $S^{(1)}$ remains the
same, while for $S^{(2)}$ we obtain
\begin{equation}
 S^{(2)}\,:\,\mathcal{H}_2\rightarrow 
 (\mathbbm{1}^{(2)}\!\!-\!\!P_0^{(2)}) 
 \mathcal{H}_2\quad\text{with}\quad
 S^{(2)\dagger}S^{(2)}=-\mathbbm{1}^{(2)}\quad\text{and}\quad
 S^{(2)}S^{(2)\dagger} = -(\mathbbm{1}^{(2)}\!\!-\!\!P_0^{(2)}),
\end{equation}
with the representation
\begin{equation}
 S^{(2)}=\sum_{n\geq 0}(-1)^n|n+1\rangle_2\langle n|_2.
\end{equation}
For an operator $S\,:\,\mathcal{H}\rightarrow
(\mathbbm{1}-P_0)\mathcal{H}$ on
$\mathcal{H}$ with the properties $SS^\dagger=\mathbbm{1}-P_0$ and 
$S^\dagger S=\mathbbm{1}$, we can use the old expression (\ref{shift1}),
since the indefiniteness of the norm does not affect the
verification of these properties. Indeed, $S$ contains only the
identity $\mathbbm{1}$, given by (\ref{eq:unew}), and the
projector $P_0^{(2)}$ on $\mathcal{H}_2$.

\paragraph{$U_+(1)$ SW monopole equations with $^t\Phi=(\phi,0)$.} 
We start from (\ref{eq:uglugl-uplus}) with $v\equiv 0$. 
In this case the SW equations read as
\begin{equation}\label{rmeq}
 \fbar{1}{1}{}+\fbar{2}{2}{}=-\sfrac{1}{4}\phi\phi^\dagger,\qquad 
 \f{1}{2}{}=0\qquad\text{and}\qquad
  D_{\bar{z}^{\bar{a}}}\phi=0,
\end{equation}
where the covariant derivative $D_{z^a}$ 
acts as $D_{z^a}\phi=-\theta_{a\bar{a}}\phi\bar{z}^{\bar{a}}+
X_{z^a}\phi$. Suppose that $\mathcal{H}_1$ is positive normed
while $\mathcal{H}_2$ is endowed with an indefinite norm. 
Using the definition (\ref{shift1}) and the ansatz  
$X_{z^a}=\theta_{a\bar{a}}S\bar{z}^{\bar{a}}S^\dagger$,
we find again (\ref{eq:commutator}).
Assuming that $\phi=\gamma P_0$ with 
$\gamma\in\mathbbm{R}$, we can solve the last equation of (\ref{rmeq}) 
if $P_0z^a=0$, i.e.,
$z^a\sim c_a^\dagger$. For $a=2$ this is satisfied because of our 
choice $\theta^{2\bar{2}}>0$ and for $a=1$ it implies 
$\theta^{1\bar{1}}<0$. The second equation of (\ref{rmeq}) is again
identically satisfied, while the first one yields the condition 
$\theta_{1\bar{1}}+\theta_{2\bar{2}}=-\sfrac{1}{4}\gamma^2$, which makes 
sense due to the different signs of $\theta_{1\bar{1}}$ and 
$\theta_{2\bar{2}}$. The nonvanishing components of the curvature
are $\fbar{a}{a}{}=\theta_{a\bar{a}}P_0$ and one can readily verify
that the current $\mathcal{J}_{\mu}$ from
(\ref{eq:current-uplus}) has no contributions to the topological charge 
(\ref{eq:tc-uplus}).  
Moreover, the computation of the topological charge gives $-1$. 
Note that the 
introduction of an indefinite norm on $\mathcal{H}_2$ 
was needed for having solutions to the equation
$\theta_{1\bar{1}}+\theta_{2\bar{2}}=-\sfrac{1}{4}\gamma^2$
on the components of $\theta_{\mu\nu}$.

\paragraph{$U_+(1)$ SW monopole equations and vortices.} 
Finally we discuss a case which is slightly different from the 
cases described above, in the sense that we fix the explicit form 
of the solutions on a two-dimensional subspace of 
$\mathbbm{R}^4_\theta$ from the very beginning. 
This eventually results in the $U_+(1)$ vortex
equations on the complementary two-dimensional subspace.

Again, let $\mathcal{H}_1$ be positive normed and $\mathcal{H}_2$ 
endowed with an indefinite norm, such that 
$[z^2,\bar{z}^{\bar{2}}]=\theta^{2\bar{2}}>0$ and 
$[c_2,c_2^\dagger]=-1$. 
Furthermore, we choose an ansatz for $X_{z^2}$ of the form 
$X_{z^2}=\theta_{2\bar{2}}\mathbbm{1}^{(1)}
\otimes S^{(2)}\bar{z}^{\bar{2}}S^{(2)\dagger}$. A short
calculation yields the result 
$\fbar{2}{2}{}=\theta_{2\bar{2}}\mathbbm{1}^{(1)}\otimes P_0^{(2)}$.
Assuming that $\mathcal{A}_{z^1}=\mathscr{A}_{z^1}\otimes P_0^{(2)}$, 
we obtain 
$\fbar{1}{1}{}=\mathscr{F}_{z^1\bar{z}^{\bar{1}}}\otimes P_0^{(2)}$. 
Similarly, we take $\phi=\psi\otimes P_0^{(2)}$. 
Then the equation $D_{\bar{z}^{\bar{2}}}\phi=0$ implies the 
condition $P_0^{(2)}z^2=0$, which is identically satisfied due 
to the relation $z^2\sim c_2^\dagger$. Moreover, the equation 
$\f{1}{2}{}=0$ is trivially satisfied. Using 
$\theta_{2\bar{2}}=-1/\theta^{2\bar{2}}$, we finally arrive at the 
equations\footnote{Note that these equations can also be obtained 
in the context of the perturbed SW equations if one chooses the 
perturbation proportional to $P_0^{(2)}$. Then there is no necessity 
to introduce an indefinite normed space.} 
\begin{subequations}\label{eq:pve}
\begin{eqnarray}
 \left\{\mathscr{F}_{z^1\bar{z}^{\bar{1}}}
 -\left(\frac{1}{\theta^{2\bar{2}}}-\frac{1}{4}
 \psi\psi^\dagger\right)\right\}\otimes P_0^{(2)} &=& 0,\\
 (\pbar{1}+\mathscr{A}_{\bar{z}^{\bar{1}}})
 \psi\otimes P_0^{(2)} &=& 0.
\end{eqnarray}
\end{subequations}
Rescaling $\psi\mapsto 2\,\tilde{\psi}/\sqrt{\theta^{2\bar{2}}}$
and introducing $\beta:=4/\theta^{2\bar{2}}$,
we obtain the equations 
\begin{equation}\label{ncveqn}
 \mathscr{F}_{z^1\bar{z}^{\bar{1}}}
 =\sfrac{\beta}{4}(1-\tilde{\psi}\tilde{\psi}^\dagger)
 \qquad\text{and}\qquad
 (\pbar{1}+\mathscr{A}_{\bar{z}^{\bar{1}}})\tilde{\psi}=0,
\end{equation}
on $\mathbbm{R}^2_\theta\subset\mathbbm{R}^4_\theta$. 
For $\beta=1$ they coincide with
the standard $U_+(1)$ vortex equations. The equations (\ref{ncveqn})
with $\theta^{1\bar{1}}=2\theta^{12}\not=0$
and their explicit solutions
were extensively discussed in the literature (see, e.g., 
\cite{Jatkar:2000ei,Bak:2000ac,Harvey:2000jb}). 
Note that the choice of the projector $P_0^{(2)}$ on $\mathcal{H}_2$ 
ensures a finite charge for this solution. To exemplify this case let
us rewrite $\mathscr{A}_{z^1}$ via $\mathscr{X}_{z^1}=\mathscr{A}_{z^1}
+\theta_{1\bar{1}}\bar{z}^{\bar{a}}$ with
$\mathscr{X}_{z^1}=\theta_{1\bar{1}}(S^{(1)})^N\bar{z}^{\bar{1}}
(S^{(1)\dagger})^N$. Moreover, suppose that $\tilde{\psi}=(S^{(1)})^N$.
Then one readily verifies that 
$\theta_{1\bar{1}}=\beta/4=1/\theta^{2\bar{2}}$.
Therefore, we end up with $\fbar{1}{1}{}=\theta_{1\bar{1}}\mathcal{P}_N$
and $\phi=\sfrac{2}{\sqrt{\theta^{2\bar{2}}}}(S^{(1)})^N\otimes P_0^{(2)}$.
Note that $\theta^{1\bar{1}}<0$. The topological charge (\ref{tc}) 
for this configuration is $Q=-N$.

To summarize, we have described solutions on
$\mathbbm{R}^2_{\theta^{12}}\times\mathbbm{R}^2_{\theta^{34}}$. 
It is also allowed to put 
$\theta^{12}$ to zero in order to have solutions on
$\mathbbm{R}^2\times\mathbbm{R}_{\theta^{34}}^2$. Then the second 
equation of (\ref{ncveqn}) reduces to 
$\mathscr{A}_{\bar{z}^{\bar{1}}}=-\pbar{1}\log\tilde{\psi}$. Assuming 
that $\tilde{\psi}=e^{(u+iv)/2}$ has zeros at points $z^1_n$ in the
complex plane, we obtain 
from the first equation of (\ref{ncveqn}) with $\beta=1$ 
(see, e.g.,~\cite{Jaffe:1980,Sergeev})
\begin{equation}
 \Delta u=e^u-1+4\pi\sum_{n=1}^N\delta^{(2)}(z^1-z^1_n,\bar{z}^{\bar{1}}
 -\bar{z}^{\bar{1}}_n),
\end{equation} 
i.e., the standard Liouville type equation on 
$\mathbbm{R}^2\cong\mathbbm{C}$. 
The moduli $z^1_n$ are position parameters of the
$N$-vortex solution on the $z^1$-plane. It is well known that
this equation exhibits regular $N$-vortex solutions 
\cite{Jaffe:1980}.

%---------------------------------------------------------------------

\subsection{Noncommutative solitons and $D$-branes}

Without refering to string theory the $U_+(1)\times U_-(1)$ and 
$U_\pm(1)$ SW monopole equations are simply understood as  
noncommutative generalizations of the Abelian SW equations. In this
section we shall discuss how one can interpret solutions to the
noncommutative SW equations as $D$-brane configurations in a stringy
context.
 
\paragraph{Brane-antibrane effective action.}
In the simplest case of type II superstrings living in the target space
$\mathbbm{R}^{9,1}$ a $Dp$-brane with a world volume 
$\mathbbm{R}^{p,1}\hookrightarrow\mathbbm{R}^{9,1}$ may be defined via a
relative map,
\begin{equation}
 \varphi\,:\,(\Sigma_2,\partial\Sigma_2)\rightarrow
 (\mathbbm{R}^{9,1},\mathbbm{R}^{p,1}),
\end{equation}
where $\Sigma_2$ is a string world sheet with boundary $\partial\Sigma_2$.
One may also consider $\overline{Dp}$-branes (= anti-$Dp$-branes) which
are $Dp$-branes with opposite orientation and Ramond-Ramond (RR) charge.

It is well known that there are stable BPS $Dp$-branes in type IIA (even $p$)
and type IIB (odd $p$) superstring theory. Besides that, it is also well
known that a system consisting of coincident $Dp$-brane and 
$\overline{Dp}$-brane 
is unstable since open strings ending on these 
branes have a tachyonic mode ($M^2<0$) in the spectrum (see, e.g., 
\cite{Witten:1998cd,Sen:1999mg,Olsen:1999xx,Ohmori:2001am,Taylor:2002uv} 
and references therein). This instability can be seen in 
the low energy effective action for the brane-antibrane pair. Namely, the
effective field theory describing light excitations of this system contains
two Abelian gauge potentials $A_\pm$ and a complex scalar $\phi$ (tachyon). 
The
latter one is associated with modes of the open string stretched between 
brane and antibrane, and is believed to be subject to a
``Mexican hat'' potential. The tachyon carries charge one under the 
diagonal $U(1)$ subgroup of the group $U_+(1)\times U_-(1)$ with the
generator $\diag{i,-i}$ corresponding to the gauge potential $A_+-A_-$.
After turning on a constant $B$-field (generating a noncommutativity
tensor $\theta$ on $\mathbbm{R}^{p,1}$ \cite{Seiberg:1999vs}) the theory 
becomes
noncommutative and the tachyon field transforms in the bi-fundamental
representation of $U_+(1)\times U_-(1)$ \cite{Harvey:2000jt}.

In perturbative string theory the first computations of the brane-antibrane
effective action were performed in \cite{Pesando:1999hm}. The resulting 
effective Lagrangian reads as
\begin{equation}\label{eq:leea}
 \mathcal{L}^{(2)}=F_{+\hat{\alpha}\hat{\beta}}\overline{F_+^{\hat{\alpha}
 \hat{\beta}}}+F_{-\hat{\alpha}\hat{\beta}}\overline{F_-^{\hat{\alpha}
 \hat{\beta}}}+(D^{\hat{\alpha}}\phi)\overline{D_{\hat{\alpha}}\phi}+
 \sfrac{1}{4}(\tau^2-\phi\bar{\phi})^2,
\end{equation}
where the covariant derivative is given by
$D_{\hat{\alpha}}\phi=\partial_{\hat{\alpha}}\phi+(A_{+\hat{\alpha}}
-A_{-\hat{\alpha}})\phi$, the overbar denotes complex conjugation, $\tau^2$ 
is a 
constant and $\hat{\alpha},\hat{\beta},\ldots=0,1,\ldots,p$. This is not 
the full effective Lagrangian since the coupling between closed string
RR fields and open strings gives an additional term 
$\mathcal{L}_{\text{CS}}$ called the Chern-Simons term 
\cite{Kennedy:1999nn}. For 
constant components of the RR fields, which we now consider, this term
does not contribute to the equations of motion, and therefore
we will not discuss it
here. Note that the terms given by (\ref{eq:cs-1}) and (\ref{eq:kchi}) 
are of such a kind.

The tachyon potential given by (\ref{eq:leea}) has the quartic form
\begin{equation}\label{eq:tp-1}
 V(\phi,\bar{\phi})=\sfrac{1}{4}(\tau^2-\phi\bar{\phi})^2.
\end{equation}
Computations in level truncated superstring field theory yield the same
result \cite{Berkovits:2000zj}. Any minimum $\phi\bar{\phi}=\tau^2$ of
the tachyon 
potential describes the closed string vacuum (tachyon condensate).
The perturbative spectrum around this vacuum is conjectured not to
contain any open string excitations. As it was discussed in
\cite{Harvey:2000jt,Harvey:2000jb,Harvey:2001yn}, in the presence
of a $B$-field background a noncommutative generalization of the 
effective theory (\ref{eq:leea}) might be of the form
\begin{equation}\label{eq:leea-nc}
 \mathcal{L}^{(2)}_{\text{nc}}=F_{+\hat{\alpha}\hat{\beta}}
 F_+^{\dagger\hat{\alpha}\hat{\beta}}+F_{-\hat{\alpha}\hat{\beta}}
 F_-^{\dagger\hat{\alpha}\hat{\beta}}+\sfrac{1}{2}\{D^{\hat{\alpha}}
 \phi,(D_{\hat{\alpha}}\phi)^\dagger\}+\sfrac{1}{8}(\tau^2-\phi
 \phi^\dagger)^2+\sfrac{1}{8}(\tau^2-\phi^\dagger\phi)^2,
\end{equation}
where $D_{\hat{\alpha}}\phi=\partial_{\hat{\alpha}}\phi+
A_{+\hat{\alpha}}\phi-\phi A_{-\hat{\alpha}}$.
Higher order corrections to the quartic tachyon potential are known
(see, e.g., \cite{Berkovits:2000hf,Ohmori:2001am}). The result is 
qualitatively similar
(``Mexican hat'' form with minima at $\phi\bar{\phi}<\infty$), and 
(\ref{eq:tp-1}) is the leading order term. Note that a quartic form of
the potential was also obtained in \cite{Gava:1997jt,Seiberg:1999vs,
Dhar:1999ax,Aganagic:2000mh} for bound states in 
$D(p-2)$-$Dp\/$
and $D(p-4)$-$Dp\/$ systems by using scattering calculations in string
theory, level truncated superstring field theory and by analyzing the
fluctuation spectrum around (noncommutative) vortex and instanton
solutions.

Quite different results have been obtained in boundary string field
theory (BSFT) (see, e.g., \cite{Kraus:2000nj,Takayanagi:2000rz} 
and references therein). There (as in \cite{Sen:1999md}) 
the Lagrangian density is proportional to the tachyon potential
itself, and the potential has the form $V\sim\exp(-\phi\bar{\phi})$ with
a ring of minima at $\phi\bar{\phi}\to\infty$. 
It is believed that the level
truncation scheme and the BSFT approach can be related by a field 
redefinition involving all components of the string field 
\cite{Kraus:2000nj}.
Various `improved versions' of the effective action of the 
brane-antibrane system have been introduced (see, e.g., 
\cite{Jones:2002si,Sen:2003tm}), and 
in the literature there is no final agreement on its form. Thus, the
action based on the Lagrangian (\ref{eq:leea}) (or (\ref{eq:leea-nc})
in the presence of a $B$-field) might be considered as an approximation
of the low energy effective action of the brane-antibrane system. 
In any case the discussed Yang-Mills-Higgs theory provides a simple
field theoretic model of the more complicated string field theory
description of brane-antibrane systems. 

\paragraph{Noncommutative vortices and ABS solitons.} 
Recall that branes and antibranes have opposite RR charge and therefore
they can annihilate into the closed string vacuum state with $A_\pm=0$ 
and $\phi\phi^\dagger=\phi^\dagger\phi=\tau^2$. 
However, instead of taking the vacuum solution,
one can choose as the ground state a (tachyon) soliton solution 
to the equations of motion for the Lagrangian (\ref{eq:leea-nc}). 
Such kinds
of solutions are interpreted as bound states of $Dp$-branes and 
$\overline{Dp}$-branes, equivalent to $D$-branes of lower dimensionality.
Here we discuss the main example of such solutions obtained via the
Atiyah-Bott-Shapiro (ABS) construction \cite{Atiyah:1964,Witten:1998cd,
Horava:1998jy,Harvey:2000te}. These ABS 
(noncommutative) solitons live in $2n\leq p$ dimensions, and for $n=1$
they coincide with noncommutative vortices on $\mathbbm{R}^2_\theta$.
For $n=2$ the ABS solitons are non-Abelian and therefore differ from
those (Abelian) SW solutions which also solve the field equations
following from (\ref{eq:leea-nc}).

To describe the ABS solitons on $\mathbbm{R}^{2n}_\theta$ we consider
a non-Abelian generalization of the Langrangian (\ref{eq:leea-nc}) 
with a $U_+(q)\times U_-(q)$ gauge group for $q=2^{n-1}$ and the 
tachyon field $\phi$ in the bi-fundamental representation $(q,\bar{q})$
of this group. This model describes two Hermitian rank $q$ vector bundles
$E_\pm\rightarrow\mathbbm{R}^{p,1}$ with connection one-forms $A_\pm$
and $\phi\in\text{Hom}(E_-,E_+)$ and is associated with a 
system of $q$ $Dp$-branes and $q$ $\overline{Dp}$-branes with common
world volume $\mathbbm{R}^{p,1}$. We assume that $2n\leq p$ and introduce
an ansatz for ABS solitons related to Clifford algebras.

Consider the Clifford algebra of the Euclidean space $(\mathbbm{R}^{2n},
\delta_{\alpha\beta})$, generated by unity and elements $\Gamma_\alpha$
such that
\begin{equation}
 \Gamma_\alpha\Gamma_\beta+\Gamma_\beta\Gamma_\alpha=-2\delta_{\alpha\beta},
\end{equation}
with $\alpha,\beta,\ldots=1,\ldots,2n$. In the $2^n\times 2^n$ matrix 
representation of this algebra the generators $\Gamma_\alpha$ can be chosen
of the form
\begin{equation}
 \Gamma_\alpha=\begin{pmatrix} 0 & \gamma_\alpha^\dagger\\
                               -\gamma_\alpha & 0\end{pmatrix},
\end{equation}
where the $\gamma_\alpha$'s are $q\times q$ gamma matrices with $q=2^{n-1}$.
In this representation the spinor space $W\cong\mathbbm{C}^{2q}$ can be
decomposed into a direct sum $W\cong W^+\oplus W^-$ of semi-spinor spaces
$W^\pm$ (the spaces of Weyl spinors). Note that for $n=1$ we have $\gamma_1=1$ 
and $\gamma_2=i$.

Considering the noncommutative deformation $\mathbbm{R}^{2n}_\theta$ of
$\mathbbm{R}^{2n}$ discussed in subsections 3.1 and 3.3, we introduce the
operators
\begin{equation}
 T=\gamma_\alpha^\dagger x^\alpha\frac{1}{\sqrt{(\gamma x)
 (\gamma x)^\dagger}}\qquad\text{and}\qquad
 T^\dagger=\frac{1}{\sqrt{(\gamma x)(\gamma x)^\dagger}}\gamma_\alpha 
 x^\alpha,
\end{equation}
where $\gamma x$ is a shorthand notation for $\gamma_\alpha x^\alpha$.
The operator $T$ defines a map,
\begin{equation}
 \hat{T}\,:\,\mathcal{H}\otimes W^-\rightarrow\mathcal{H}\otimes W^+.
\end{equation}
Here $\mathcal{H}$ is the Hilbert space realized as a representation
space of $n$ oscillators defined by formulas similiar to 
(\ref{eq:cao}) and (\ref{eq:cao-z}), and $W^\pm$ are the semi-spinor 
spaces introduced above. In matrix realization $\hat{T}$ looks as
\begin{equation}
 \hat{T}=\begin{pmatrix} 0 & T\\ 0 & 0\end{pmatrix},
\end{equation}
and hence, following the authors of \cite{Witten:1998cd,Witten:2000cn,
Harvey:2000te,Harvey:2001yn}, 
we will not distinguish $T$ and $\hat{T}$ in the sequel.

It is not difficult to see that
\begin{equation}
 T^\dagger T=\mathbbm{1}_q\qquad\text{and}\qquad
 TT^\dagger=\mathbbm{1}_q-\mathcal{P}_0,
\end{equation}
where $\mathcal{P}_0$ is the projector onto the kernel of $T^\dagger$.
This state is the tensor product of the oscillator ground state with
itself and
the lowest weight spinor of $SO(2n)$ (the fermion ground state). Also,
by introducing $T_N:=T^N$ with $T_1\equiv T$, we have
\begin{equation}
 T_N^\dagger T_N=\mathbbm{1}_q\qquad\text{and}\qquad
 T_NT_N^\dagger=\mathbbm{1}_q-\mathcal{P}_{N-1},
\end{equation}
where $\mathcal{P}_{N-1}$ is the projector onto the kernel of 
$T^\dagger_N$, an $N$-dimensional subspace in $\mathcal{H}\otimes W$.
One can show that
\begin{equation}
 \dim\ker T_N=0\qquad\text{and}\qquad
 \dim\text{coker}\,T_N:=\dim\ker T_N^\dagger=N,
\end{equation} 
and therefore the index of $T_N$ is given by
\begin{equation}
 \text{ind}\,T_N:=\dim\ker T_N-\dim\text{coker}\,T_N=-N.
\end{equation}
The operators $T_N$ and $T_N^\dagger$ are Toeplitz operators.

Now we reduce the equations of motion for the Lagrangian 
(\ref{eq:leea-nc}) to the space $\mathbbm{R}^{2n}_\theta\hookrightarrow
\mathbbm{R}^{p+1}_\theta$ by assuming that all fields depend only on
$x^\alpha$ and by taking $A_{\pm\alpha}$ as the only nonvanishing 
components of the gauge potentials $A_\pm$. To solve the reduced field
equations we consider $T_N$ as $q\times q$ matrices with operator
entries and introduce the ansatz, 
\begin{equation}\label{eq:sol-leea-nc}
 A_{+z^a}=-(A_{+\bar{z}^{\bar{a}}})^\dagger=\tau\theta_{a\bar{a}}
 (T_{N_+}\bar{z}^{\bar{a}}T_{N_+}^\dagger-\bar{z}^{\bar{a}}),\qquad
 A_{-z^a}=0=A_{-\bar{z}^{\bar{a}}}
 \qquad\text{and}\qquad\phi=\tau T_{N_+},
\end{equation}  
which solves the equation of motion for (\ref{eq:leea-nc}) 
\cite{Harvey:2000jb,Harvey:2001yn}.
More general solutions can also be constructed. For $n=1$ the solution
(\ref{eq:sol-leea-nc}) describes $N_+$ vortices on 
$\mathbbm{R}^2_\theta$.
In the case $n=2$ it gives a solution for the
$U_+(2)\times U_-(2)$ Yang-Mills-Higgs model on $\mathbbm{R}^4_\theta$.
Some noncommutative SW solutions (e.g., (\ref{eq:ax}), (\ref{eq:ap}) 
with $N_-=0$)
solve the above equations as well but for the gauge group
$U_+(1)\times U_-(1)$. These solutions can therefore be regarded as 
a new kind of tachyon solitons.

\paragraph{Noncommutative SW solitons.}
Note that the SW equations (\ref{eq:uglugl}) with $v_+=v_-=\tau^2=
\text{const}$ coincide with the first order BPS equations for the
Lagrangian (\ref{eq:leea-nc}), and therefore their solutions also satisfy 
the equations of motion for (\ref{eq:leea-nc}). In particular,
the solution given by (\ref{eq:ax-ap}) with $N_+\geq 1$ and 
$N_-=0$ is such a solution. Following the general logic of Sen's proposal,
it is natural to interpret this solution as a configuration of 
$N_+$ stable $D(p-4)$-branes,
corresponding to the topologically stable codimension four SW soliton on a 
$Dp-\overline{Dp}$ brane pair.
 
The more general configuration described by (\ref{eq:ax-ap})
with $N_{\pm}\geq 1$ can be interpreted in two different ways. First, we
may again choose $v_+=v_-=\tau^2=\text{const}$. Then one can show that this
configuration satisfies the second order equations of motion for the 
Lagrangian (\ref{eq:leea-nc}) but does not satisfy the first order 
equations (\ref{eq:uglugl}) with constant $v_\pm$. This is natural since
this solution corresponds to a system of $N_+$ $D(p-4)$-branes and $N_-$
$\overline{D(p-4)}$-branes which is not a BPS configuration from the point
of view of the Lagrangian (\ref{eq:leea-nc}). Second, one may consider
the situation where $v_+$ depends on $\phi\phi^\dagger$ and $v_-$ on 
$\phi^\dagger\phi$ and take as the low energy effective action the sum
of the functional (\ref{eq:echi}) and the Chern-Simons term 
(\ref{eq:kchi-1}), where
the latter contributes to the equations of motion for nonconstant
$v_\pm$. Then for proper choices of $v_\pm$ one can obtain configurations
which satisfy the noncommutative SW equations. For instance, the choice
\begin{equation}\label{eq:vpm}
 v_+=\tau^2\qquad\text{and}\qquad v_-=-\tau^2+2\phi^\dagger\phi
\end{equation}
corresponds to the same tachyon potential $V\sim(\tau^2-\phi\phi^\dagger)^2+
(\tau^2-\phi^\dagger\phi)^2$ as in (\ref{eq:leea-nc}). However, for the
above choice of $v_\pm$ the Chern-Simons term (\ref{eq:kchi-1}) becomes 
nontrivial
and contributes to the equations (\ref{eq:uglugl}) which are the BPS
equations for the action $E_\chi+16\pi^2 K_\chi$ with $E_\chi$ and $K_\chi$
determined by (\ref{eq:e-k-chi}). Therefore, for $v_\pm$ given by 
(\ref{eq:vpm})
the configuration (\ref{eq:ax-ap}) with $N_\pm\geq 1$ is a 
solution to the SW equations (\ref{eq:uglugl}). So, in both cases the
configuration (\ref{eq:ax-ap}) may be interpreted as $N_+$
$D(p-4)$-branes and $N_-$ $\overline{D(p-4)}$-branes. Other solutions
to the noncommutative SW equations can be analyzed similarly.
  
%---------------------------------------------------------------------

\section{Concluding remarks}

In this paper we have discussed different noncommutative deformations
of the (perturbed) SW monopole equations on Euclidean 
four-dimensional space. Namely, starting from properly deformed $U(2)$
self-duality type YM equations in eight dimensions, we performed a
reduction to $U(2)$ noncommutative SW equations on 
$\mathbbm{R}^4_\theta$ with the matter field in the
adjoint representation of the gauge group. We then concentrated on the
 $U_+(1)\times U_-(1)\subset U(2)$ noncommutative SW equations with the 
matter in the bi-fundamental representation of 
$U_+(1)\times U_-(1)$. Perturbed versions of these
equations have also been discussed. Then, by considering the matter field
$\Phi$ as an element of a right or left $\mathbbm{R}^4_\theta$-module,
we have introduced the (perturbed) $U_+(1)$ and $U_-(1)$ SW equations.
The commutative limits of all these three Abelian cases are identical to
the standard (perturbed) Abelian SW equations on $\mathbbm{R}^4$.  
In summary we may write down the following diagram:

{\footnotesize
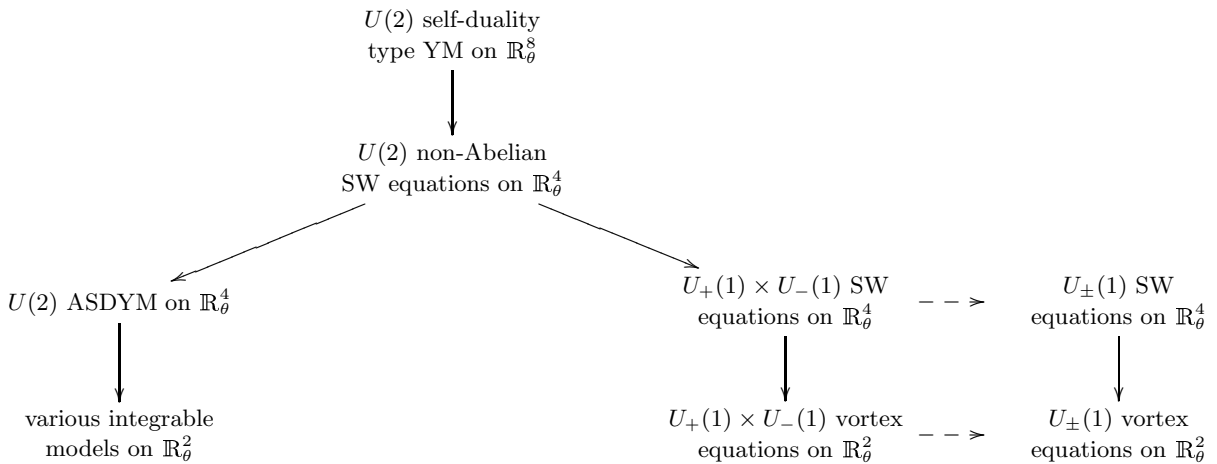
\begin{center}
\xymatrix{
 & \txt<8pc>{$U(2)$ self-duality type YM on $\mathbbm{R}^8_\theta$}
 \ar[d] \\
 & \txt<8pc>{$U(2)$ non-Abelian SW equations on $\mathbbm{R}^4_\theta$} 
 \ar[dl]\ar[dr] \\
 \txt<8pc>{$U(2)$ ASDYM on $\mathbbm{R}^4_\theta$} \ar[d] &  &
 \txt<8pc>{$U_+(1)\times U_-(1)$ SW equations on $\mathbbm{R}^4_\theta$}
 \ar[d] \ar@{-->}[r]   & \txt<8pc>{$U_\pm(1)$ SW equations on 
 $\mathbbm{R}^4_\theta$}\ar[d]\\
 \txt<8pc>{various integrable models on $\mathbbm{R}^{2}_\theta$}  
 & &\txt<8pc>{$U_+(1)\times U_-(1)$ vortex equations on 
 $\mathbbm{R}^2_\theta$} \ar@{-->}[r]  & \txt<8pc>{$U_\pm(1)$ vortex 
 equations on $\mathbbm{R}^2_\theta$}
}
\end{center}
}
\noindent
This picture shows the connection between the theories discussed in this
paper. Note that the noncommutative vortex equations in two dimensions
can easily be obtained via dimensional reduction from the noncommutative
SW equations with properly chosen perturbations.

It has been shown that $\mathbbm{R}^4_\theta$
supports regular finite-action solutions to the SW equations even if
there are no such solutions in the commutative case. This
is a well known phenomenon related to the fact that, due to the 
noncommutativity tensor $\theta$, an additional length scale enters the
theory. We have constructed explicit solutions to the 
$U_+(1)\times U_-(1)$ and $U_\pm(1)$ noncommutative SW equations and
interpreted them as $D$-brane configurations in type II superstring
theory. It would be illuminating to generalize the present results
to non-Abelian noncommutative SW theory and discuss the latter's
relation to superstring theory. 

%---------------------------------------------------------------------

\subsection*{Acknowledgments}

We are grateful to O.Lechtenfeld for useful comments. The work of
A.D.P. and A.G.S. was supported in part by the grants
DFG 436 RUS 113/669 and RFBR 02-02-04002. 
M.W. is grateful to the Studienstiftung des deutschen Volkes for 
financial support.
This work was done within the framework of the DFG priority 
program in string theory and supported by the grant Le 838/7.

\setcounter{section}{0}
\renewcommand{\thesection}{\Alph{section}}

%---------------------------------------------------------------------

\section{Noncommutative Bogomolny transformation}

In order to derive (\ref{bog}) we need the trivial identities
\begin{equation}\label{ti}
 (D_\mu\phi)\psi^\dagger=-\phi(D_\mu\psi)^\dagger+
   \nabla_{+\mu}(\phi\psi^\dagger)
 \qquad\text{and}\qquad
 (D_\mu\phi)^\dagger\psi=-\phi^\dagger(D_\mu\psi)+
 \nabla_{-\mu}(\phi^\dagger\psi),
\end{equation}
where the covariant derivatives
$\nabla_{\pm\mu}$ are given by
(\ref{nabla1}). Furthermore, we have
\begin{equation}\label{ti2}
 [D_\mu,D_\nu]\phi=F_{+\mu\nu}\,\phi-\phi\, F_{-\mu\nu}.
\end{equation}
Let us consider
\begin{align}
 \hspace*{-0.4cm}4|F^+_{A_+}+\chi^+_+-\sigma^+(\Phi\otimes\Phi^\dagger)_0|^2
  = &-2(F^+_{+\mu\nu}+\chi^+_{+\mu\nu}-\sigma^+_{\mu\nu}
    (\Phi\otimes\Phi^\dagger)_0)^2\notag\\
  = & -2(F_{+\mu\nu}^+)^2-2\{F^+_{+\mu\nu},\chi^+_{+\mu\nu}\}
      +2\{F^+_{+\mu\nu},\sigma^+_{\mu\nu}(\Phi\otimes\Phi^\dagger)_0\}
      \notag\\
    & \qquad
      -2(\chi^+_{+\mu\nu}-\sigma^+_{\mu\nu}(\Phi\otimes\Phi^\dagger)_0)^2.
\end{align}
This expression contains
\begin{subequations}
\begin{eqnarray}
 (F_{+\mu\nu}^+)^2 &=& \sfrac{1}{4}(F_{+\mu\nu}+{*F_{+\mu\nu}})^2=
 \sfrac{1}{2}(F_{+\mu\nu})^2+\sfrac{1}{4}\{F_{+\mu\nu},{*F_{+\mu\nu}}\},\\
 \{F^+_{+\mu\nu},\sigma^+_{\mu\nu}(\Phi\otimes\Phi^\dagger)_0\} &=&
 \sfrac{i}{2}\{F_{+12}^+,\phi_1\phi_1^\dagger-\phi_2\phi_2^\dagger\}-
 \sfrac{1}{2}\{F_{+13}^+,\phi_2\phi_1^\dagger-\phi_1\phi_2^\dagger\}\notag\\
  &&\qquad +\,\,\sfrac{i}{2}\{F_{+14}^+,\phi_2\phi_1^\dagger
      +\phi_1\phi_2^\dagger\},\\
  -(\chi^+_{+\mu\nu}-\sigma^+_{\mu\nu}(\Phi\otimes\Phi^\dagger)_0)^2&=&
  2|\chi^+_+-\sigma^+(\Phi\otimes\Phi^\dagger)_0|^2.
\end{eqnarray}
\end{subequations}
Similarly,
\begin{align}
 \hspace*{-0.4cm}4|F^+_{A_-}+\chi^+_--\sigma^+(\Phi^*\otimes(\Phi^*)^
\dagger)_0|^2
  = &-2(F^+_{-\mu\nu}+\chi^+_{-\mu\nu}-\sigma^+_{\mu\nu}
    (\Phi^*\otimes(\Phi^*)^\dagger)_0)^2\notag\\
  = & -2(F_{-\mu\nu}^+)^2-2\{F^+_{-\mu\nu},\chi^+_{-\mu\nu}\}
      +2\{F^+_{-\mu\nu},\sigma^+_{\mu\nu}(\Phi^*\otimes(\Phi^*)^\dagger)_0\}
      \notag\\
    & \qquad
      -2(\chi^+_{-\mu\nu}-\sigma^+_{\mu\nu}(\Phi^*\otimes(\Phi^*)^
\dagger)_0)^2
\end{align}
with
\begin{subequations}
\begin{eqnarray}
 (F_{-\mu\nu}^+)^2 &=& \sfrac{1}{2}(F_{-\mu\nu})^2+
  \sfrac{1}{4}\{F_{-\mu\nu},{*F_{-\mu\nu}}\},\\
 \{F^+_{-\mu\nu},\sigma^+_{\mu\nu}(\Phi^*\otimes(\Phi^*)^\dagger)_0\} &=&
 -\sfrac{i}{2}\{F_{-12}^+,\phi_1^\dagger\phi_1-\phi_2^\dagger\phi_2\}+
 \sfrac{1}{2}\{F_{-13}^+,\phi_1^\dagger\phi_2-\phi_2^\dagger\phi_1\}\notag\\
  &&\qquad -\,\,\sfrac{i}{2}\{F_{-14}^+,\phi_1^\dagger\phi_2
      +\phi_2^\dagger\phi_1\},\\
  -(\chi^+_{-\mu\nu}-\sigma^+_{\mu\nu}(\Phi^*\otimes(\Phi^*)^\dagger)_0)^2&=&
  2|\chi^+_--\sigma^+(\Phi^*\otimes(\Phi^*)^\dagger)_0|^2.
\end{eqnarray}
\end{subequations}

A lengthy but straightforward 
calculation exploiting (\ref{ti}) and (\ref{ti2})
yields for $\sfrac{1}{2}|\mathcal{D}_{A_+,A_-}\Phi|^2$ the 
expression
\begin{eqnarray}
 \sfrac{1}{2}|\mathcal{D}_{A_+,A_-}\Phi|^2
    &=&\sfrac{1}{2}|iD_3\phi_1-D_4\phi_1+iD_1\phi_2+D_2\phi_2|^2+
       \sfrac{1}{2}|iD_1\phi_1-D_2\phi_1-iD_3\phi_2-D_4\phi_2|^2
    \notag\\
  &=&-\,\,\sfrac{i}{2}\{F_{+12}^+,\phi_1\phi_1^\dagger
       -\phi_2\phi_2^\dagger\}+\sfrac{1}{2}\{F_{+13}^+,
       \phi_2\phi_1^\dagger-\phi_1\phi_2^\dagger\}
  -\sfrac{i}{2}\{F_{+14}^+,\phi_2\phi_1^\dagger
      +\phi_1\phi_2^\dagger\}\notag\\
 &&\hspace*{1in}+\,\,\sfrac{1}{2}|D_{A_+,A_-}\Phi|^2-
 \nabla_{+\mu}\mathcal{J}_{+\mu}+C_+,
\end{eqnarray}
where
\begin{subequations}
\begin{equation}
 \mathcal{J}_{+\mu}=\sfrac{i}{4}(\epsilon_{\mu\nu12}+
         \epsilon_{\mu\nu34})\mathcal{J}_{+\nu}^{(1)}
                + \sfrac{1}{4}(\epsilon_{\mu\nu31}+
         \epsilon_{\mu\nu24})\mathcal{J}_{+\nu}^{(2)}
                + \sfrac{i}{4}(\epsilon_{\mu\nu23}+
         \epsilon_{\mu\nu14})\mathcal{J}_{+\nu}^{(3)}
\end{equation}
and
\begin{eqnarray}
 \mathcal{J}_{+\nu}^{(1)}&=&\phi_1(D_\nu\phi_1)^\dagger-
                         (D_\nu\phi_1)\phi_1^\dagger-
                         \phi_2(D_\nu\phi_2)^\dagger+
                         (D_\nu\phi_2)\phi_2^\dagger,\\
 \mathcal{J}_{+\nu}^{(2)}&=&-\phi_1(D_\nu\phi_2)^\dagger+
                         (D_\nu\phi_1)\phi_2^\dagger+
                         \phi_2(D_\nu\phi_1)^\dagger-
                         (D_\nu\phi_2)\phi_1^\dagger,\\
 \mathcal{J}_{+\nu}^{(3)}&=&\phi_2(D_\nu\phi_1)^\dagger-
                         (D_\nu\phi_2)\phi_1^\dagger+
                         \phi_1(D_\nu\phi_2)^\dagger-
                         (D_\nu\phi_1)\phi_2^\dagger.
\end{eqnarray}
\end{subequations}
The term $C_+$ is given by
\begin{equation}
 C_+=-i\phi_2F_{-12}^+\phi_2^\dagger+i\phi_1F_{-12}^+\phi_1^\dagger
 -\phi_2F_{-13}^+\phi_1^\dagger+i\phi_2F_{-14}^+\phi_1^\dagger
 +i\phi_1F_{-14}^+\phi_2^\dagger+\phi_1F_{-13}^+\phi_2^\dagger.
\end{equation}
In a similar manner, we also have
\begin{equation}
 \sfrac{1}{2}|(\mathcal{D}_{A_+,A_-}\Phi)^\dagger|^2
    =\sfrac{i}{2}\{F_{-12}^+,\phi_1^\dagger \phi_1
       -\phi_2^\dagger\phi_2\}-\sfrac{1}{2}\{F_{-13}^+,
       \phi_1^\dagger\phi_2-\phi_2^\dagger\phi_1\}
  +\sfrac{i}{2}\{F_{-14}^+,\phi_1^\dagger\phi_2
      +\phi_2^\dagger\phi_1\}\notag
\end{equation}
\begin{equation}
 +\sfrac{1}{2}|(D_{A_+,A_-}\Phi)^\dagger|^2-
 \nabla_{-\mu}\mathcal{J}_{-\mu}+C_-,
\end{equation}
where
\begin{subequations}
\begin{equation}
 \mathcal{J}_{-\mu}=-\sfrac{i}{4}(\epsilon_{\mu\nu12}+
         \epsilon_{\mu\nu34})\mathcal{J}_{-\nu}^{(1)}
                - \sfrac{1}{4}(\epsilon_{\mu\nu31}+
         \epsilon_{\mu\nu24})\mathcal{J}_{-\nu}^{(2)}
                - \sfrac{i}{4}(\epsilon_{\mu\nu23}+
         \epsilon_{\mu\nu14})\mathcal{J}_{-\nu}^{(3)}
\end{equation}
and
\begin{eqnarray}
 \mathcal{J}_{-\nu}^{(1)}&=&\phi_1^\dagger(D_\mu\phi_1)-
                         (D_\mu\phi_1)^\dagger\phi_1
                         -\phi_2^\dagger(D_\mu\phi_2)+
                         (D_\mu\phi_2)^\dagger\phi_2,\\
 \mathcal{J}_{-\nu}^{(2)}&=&-\phi_2^\dagger(D_\mu\phi_1)+
                          (D_\mu\phi_2)^\dagger\phi_1+
                          \phi_1^\dagger(D_\mu\phi_2)-
                          (D_\mu\phi_1)^\dagger\phi_2,\\
 \mathcal{J}_{-\nu}^{(3)}&=&\phi_1^\dagger(D_\mu\phi_2)-
                          (D_\mu\phi_1)^\dagger\phi_2+
                          \phi_2^\dagger(D_\mu\phi_1)-
                          (D_\mu\phi_2)^\dagger\phi_1.
\end{eqnarray}
\end{subequations}
The term $C_-$ is 
\begin{equation}
 C_-=i\phi_2^\dagger F_{+12}^+\phi_2-i\phi_1^\dagger F_{+12}^+
 \phi_1-\phi_2^\dagger F_{+13}^+\phi_1-i\phi_2^\dagger F_{+14}^+\phi_1
 -i\phi_1^\dagger F_{+14}\phi_2+\phi_1^\dagger F_{+13}\phi_2.
\end{equation}
Therefore, we discover that
\begin{align}
 & \sfrac{1}{2}|\mathcal{D}_{A_+,A_-}\Phi|^2+ 
 \sfrac{1}{2}|(\mathcal{D}_{A_+,A_-}\Phi)^\dagger|^2+
 4|F^+_{A_+}+\chi^+_+-\sigma^+(\Phi\otimes\Phi^\dagger)_0|^2+
 4|F^+_{A_-}+\chi^+_--\sigma^+(\Phi^*\otimes(\Phi^*)^\dagger)_0|^2\notag\\
 &= \sfrac{1}{2}|D_{A_+,A_-}\Phi|^2+ \sfrac{1}{2}|(D_{A_+,A_-}\Phi)^\dagger|^2
 +2|F_{A_+}|^2+2|F_{A_-}|^2
 +4|\chi^+_+-\sigma^+(\Phi\otimes\Phi^\dagger)_0|^2\notag\\
 &\qquad +4|\chi^+_--\sigma^+(\Phi^*\otimes(\Phi^*)^\dagger)_0|^2
 -\sfrac{1}{2}\{F_{+\mu\nu},{*F_{+\mu\nu}}\}-
 2\{F_{+\mu\nu}^+,\chi^+_{+\mu\nu}\}-
     \nabla_{+\mu}\mathcal{J}_{+\mu}\notag\\
 &\qquad -\sfrac{1}{2}\{F_{-\mu\nu},{*F_{-\mu\nu}}\}-
 2\{F_{-\mu\nu}^+,\chi^+_{-\mu\nu}\}-
     \nabla_{-\mu}\mathcal{J}_{-\mu}\notag\\
 &\qquad + \sfrac{i}{2}\{F_{+12}^+,\phi_1\phi_1^\dagger
 -\phi_2\phi_2^\dagger\}-\sfrac{1}{2}\{F_{+13}^+,
       \phi_2\phi_1^\dagger-\phi_1\phi_2^\dagger\}
  +\,\,\sfrac{i}{2}\{F_{+14}^+,\phi_2\phi_1^\dagger
      +\phi_1\phi_2^\dagger\}+C_+\notag\\
 &\qquad - \sfrac{i}{2}\{F_{-12}^+,\phi_1^\dagger\phi_1
 -\phi_2^\dagger\phi_2\}+\sfrac{1}{2}\{F_{-13}^+,
       \phi_1^\dagger\phi_2-\phi_2^\dagger\phi_1\}
  -\,\,\sfrac{i}{2}\{F_{-14}^+,\phi_1^\dagger\phi_2
      +\phi_2^\dagger\phi_1\}+C_-.\notag
\end{align}
Now suppose that all operators entering this formula 
are of proper trace-class, e.g.,
 $|\Tr{}\phi_{1,2}|<\infty$ and 
$|\Tr{}F_{\pm\mu\nu}|<\infty$. Then
\begin{equation}
\begin{split}
 \Tr{}&\left\{\sfrac{i}{2}\{F_{+12}^+,\phi_1\phi_1^\dagger
 -\phi_2\phi_2^\dagger\}-\sfrac{1}{2}\{F_{+13}^+,
       \phi_2\phi_1^\dagger-\phi_1\phi_2^\dagger\}
  +\,\,\sfrac{i}{2}\{F_{+14}^+,\phi_2\phi_1^\dagger
      +\phi_1\phi_2^\dagger\}+C_+\right.\\
 &\left.\quad - \sfrac{i}{2}\{F_{-12}^+,\phi_1^\dagger\phi_1
 -\phi_2^\dagger\phi_2\}+\sfrac{1}{2}\{F_{-13}^+,
       \phi_1^\dagger\phi_2-\phi_2^\dagger\phi_1\}
  -\,\,\sfrac{i}{2}\{F_{-14}^+,\phi_1^\dagger\phi_2
      +\phi_2^\dagger\phi_1\}+C_-\right\}=0,\label{eq:0}
\end{split}
\end{equation}
since we can use the invariance of the trace under cyclic permutations. 
In fact, one can easily check that for each term in (\ref{eq:0}) 
there exists a corresponding
term having the opposite sign so that the trace is indeed zero.

Finally we obtain
\begin{align}
 &\Tr{}\left\{\sfrac{1}{2}|\mathcal{D}_{A_+,A_-}\Phi|^2+ 
 \sfrac{1}{2}|(\mathcal{D}_{A_+,A_-}\Phi)^\dagger|^2+
 4|F^+_{A_+}+\chi^+_+-\sigma^+(\Phi\otimes\Phi^\dagger)_0|^2\right.\notag\\
 &\left.\hspace{2.5in}+\,\,
 4|F^+_{A_-}+\chi^+_--\sigma^+(\Phi^*\otimes(\Phi^*)^\dagger)_0|^2\right\}
\notag\\
 &\hspace*{1cm}=\Tr{}\left\{ \sfrac{1}{2}|D_{A_+,A_-}\Phi|^2+ 
 \sfrac{1}{2}|(D_{A_+,A_-}\Phi)^\dagger|^2+2|F_{A_+}|^2+2|F_{A_-}|^2
 +4|\chi^+_+-\sigma^+(\Phi\otimes\Phi^\dagger)_0|^2\right.\notag\\
 &\left.\hspace*{5cm} +\,\,4|\chi^+_--\sigma^+(\Phi^*\otimes
 (\Phi^*)^\dagger)_0|^2\right\}
 -\Tr{}\mathcal{T}+\frac{16\pi^2}{|\text{Pf}(2\pi\theta)|}K_\chi,
\end{align}
which is the desired result. Note that the choice $\phi_1=\phi$ and
$\phi_2=0$ yields the expressions (\ref{rc1}) and (\ref{rc1-2}) for the 
currents
$\mathcal{J}_{\pm\mu}$. 

%---------------------------------------------------------------------

\end{document}